# Solving Crystal Structures by Carrying Out the Calculation of the Single-Atom R1 Method in a Lottery Mode

Authors

**Xiaodong Zhang[a]***

[a]Chemistry Department, Tulane University, 6400 Freret Street, New Orleans, Louisiana, 70118, United States

Correspondence email: xzhang2@tulane.edu

**Synopsis** A lottery scheme of carrying out the single-atom R1 (sR1) calculation has been designed for solving crystal structures.

**Abstract** As originally designed [Zhang & Donahue (2024), *Acta Cryst.* A**80**, 237-248.], after one cycle of calculation, the single-atom R1 (sR1) method required a user to intelligently determine a partial structure to start the next cycle. In this paper, a lottery scheme has been designed to randomly split a parent model into two child models. This allows the calculation to be carried out in a care-free manner. By chance, one child model may have higher amounts of "good" atoms than the parent model. Thus, its expansion in the next cycle favors an improved model. These "lucky" results are carried onto the next cycles, while "unlucky" results in which no improvements occur are discarded. Furthermore, unchanged models are carried onto the next cycles in those "unlucky" occasions. On average a child model has the same fraction of "good" atoms as the parent. Only a substantial statistical fluctuation results in appreciable deviation. This lottery scheme works because such fluctuations do happen. Indeed, test applications with the computing power accessible by the author have demonstrated that the designed scheme can drive an sR1 calculation to or close to reaching a correct structure solution.

## 1. Introduction

In X-ray crystallography, many methods for solving crystal structures rely on phasing (Giacovazzo, 2014). The modern driving force behind many phasing methods is a dual space recycling process, utilizing reciprocal space phase refinement and/or real space electron density modifications (Weeks *et al.*, 1993; Miller *et al.*, 1993; DeTitta *et al.*, 1994). One way of phasing in macromolecular crystallography is the molecular replacement (MR) method (Rossmann & Blow, 1962; Rossmann *et al.*, 1964; Crowther, 1972; Rossmann, 1972, 1990; Bricogne, 1992; Read, 2001; McCoy, 2004; Read & McCoy, 2016; McCoy *et al.*, 2017; Caliandro *et al.,* 2009; Burla *et al.*, 2020). The key step of an MR calculation is using a model-searching technique to orient and translationally position a known model (or models) in an unknown structure. Originally, Patterson targets (rotation function, Rossmann & Blow, 1962; translation function, Rossmann *et al.*, 1964) were used for this task. Currently, the log-likelihood-gain on intensities (LLGI), which is the sum of log-likelihoods for individual reflections minus the log-likelihood of an uninformative model (McCoy *et al.*, 2007; Read & McCoy, 2016), is used as it is regarded as the most sensitive target for model searching. In early days of the MR method, the traditional R1 was occasionally used as a target with some success in translational search (Eventoff *et al.*, 1975; Bott & Sarma, 1976). Today, the traditional R1 is regarded as a poor target for model searching (McCoy, 2017). But a recent paper (Zhang & Donahue, 2024) yielded rather surprising results. In that publication, an approximate R1, called the single-atom R1 (sR1), was used to solve small molecule crystal structures by locating missing atoms of a structure one at a time. Essentially, the sR1 was used as a target for single atom search, and it was very successful in this regard. In fact, the whole structure of each tested case was directly solved by the sR1 method, with no need of phasing.

The calculation of the sR1 method is carried out in cycles, in which each cycle expands a starting partial model into a tentative full model (Zhang & Donahue, 2024). As originally designed, a transition from one cycle to the next requires a user to intelligently determine the next starting partial model. This is done either by selecting recognizable fragments, or by manually deleting ghost atoms. The aim of this paper is to relax this requirement by designing a lottery scheme that automatically generates starting partial models. Such a scheme has been designed with its effectiveness verified by test calculations. The paper will also discuss implications of this novel design.

## 2. Recap of the nature of the approximation in defining the single-atom R1 (sR1)

The sR1 is derived from the traditional R1 by approximation. Readers should refer to the previous paper (Zhang & Donahue, 2024) for the explicit formulas. However, it is instructive to recap the main ideas leading to this approximation. The goal is to solve a structure with $N$ atoms in its unit cell. The starting point is a partial model with atoms 1 to $i-1$ already being located. The coordinates of the atoms from $i$ to $N$ are still undetermined. Usually, it is a daunting task to optimize the coordinates of all these undetermined atoms together. It is much easier to optimize a single atom or a group of atoms (for example, a fragment with a known structure) rather than all the undetermined atoms. Let atoms $i$ to $j$ be the single atom (if $i = j$) or the group of atoms (if $i < j$) that one wishes to focus on. The atoms from $j + 1$ to $N$ are the other remaining undetermined atoms, whose fractional coordinates one wishes to get rid of from the formula of the R1. Thus, an approximation is made by deleting terms containing these coordinates from the R1 formula. However, the resulting approximate R1 is not completely unrelated to these other undetermined atoms. It is no longer related to the coordinates of these atoms, but it is still related to their atomic scattering factors through a term of the sum of the square of the atomic scattering factors of these atoms. This term is retained for the purpose of calculating the approximate R1 as accurately as possible. This approximate R1 is called the single-atom R1 (sR1) if $i = j$, or the partial-structure R1 (pR1) if $i < j$. In the sR1, atoms 1 to $j - 1$ are the ones that have already been located; atom $j$ is the only single undetermined atom whose coordinates one wants to determine. In the pR1, atoms 1 to $i - 1$ are the ones already being located, and atoms $i$ to $j$ belong to a fragment whose orientation and translational position one wants to determine.

The atomic scattering factors, which are calculated with an isotropic displacement parameter U=0 $Å^2$, are used for calculating the approximate R1. These atomic scattering factors are related to the underlying atomic electron density of spherical shapes. One intuition for understanding the sR1 method is to notice that the centers of atomic electron density spheres of atoms 1 to $j - 1$ are fixed in the unit cell. On the other hand, the center of atom $j$ is moving around in the cell, and one is trying to see which $j$ location can best account the diffraction effects. However, one should realize that because the dependence of the sR1 on the coordinates of atoms $j + 1$ to $N$ has been removed while the dependence on the atomic scattering factors of these atoms is partially retained, a correspondence to the spheres of the atomic electron density for these atoms is lost. This subtlety, in addition to the fact that the sR1 method does not explicitly calculate an electron density, makes the method not even implicitly related to an electron density of some partial structure.

## 3. Improvements to the implementation of the sR1 method

The basic implementation of the sR1 method remains the same as reported previously (Zhang & Donahue, 2024). However, two improvements have been made. First, instead of directly feeding raw reflection intensities to the calculation, sharpened intensities are used. A sharpened intensity is calculated as the product of multiplying a raw intensity ${F_0}^2(hkl)$ with exp $(2Bs^2)$, where $s = sin\theta/\lambda$. The parameter $B$ is determined by the Wilson method. Second, the precision of the coordinates of the single atoms are refined to 0.2 Å, instead of previous 0.001 Å. This cuts the calculation time while still yielding good solutions.

## 4. Designing a lottery scheme for transition from one sR1 calculation cycle to the next

The idea of playing lottery in the sR1 calculation was inspired by one published work (Burla *et al.*, 2018). In that work, a diagonal least-squares technique was used to produce *ab initio* phasing of small molecules. During the calculation, poorly located atoms were identified and replaced with new atoms that were randomly positioned.

Each cycle of an sR1 calculation expands a starting partial model into a tentative full model. To start the next cycle, a new starting partial model needs to be determined. In the original design (Zhang & Donahue, 2024), a user determines this new starting partial model either by selecting a recognizable fragment (or fragments) or by deleting ghost atoms from the tentative full model. Therefore, that design can be regarded as performing the calculation in an intelligent mode.

Ideally, the new starting partial model should only include those atoms that have been correctly located. The question is how to identify these correctly located atoms. Or, conversely, how does one identify those more-likely poorly located atoms? Burla *et al.* (2018) were able to identify poorly located atoms by checking their anisotropic displacement parameters in a diagonal least-squares calculation. But inserting a least-squares calculation slows down the sR1 method. A better alternative is to borrow another idea from another published work (Kinneging & de Graaff, 1984) in which some R1 calculation was used to identify poorly located atoms. In the sR1 calculation, when a new atom is added to the growing model, the approximate R1 usually drops. Intuitively, how much the approximate R1 drops measures how well the newly added atom contributes to accounting the diffraction effects. Reciprocally, if the addition of an atom causes the approximate R1 to rise, it is very likely this atom is incorrectly positioned. In fact, this has been verified by applying the sR1 method to about 224 datasets (a summary of these test calculations is given in the supporting information). Often, such atoms are ghost atoms. Only very rarely does adding correct atoms make the approximate R1 rise. Therefore, in the new calculation scheme, all the atoms that cause the approximate R1 to rise will be excluded when generating a new starting partial model. It should be acknowledged that a reviewer compared this process of rejecting atoms to iterative peaklist optimization where peaks are eliminated one by one to maximize CC (Sheldrick & Gould, 1995).

After excluding these very likely bad atoms, the remaining atoms are not all correctly located. But trying to figure out further which atoms are more likely correctly located is difficult. Therefore, a lottery must be played. A parent model is formed with the remaining atoms and is randomly split into two child models. On average a child model contains the same fraction of "good" atoms as the parent. Only a substantial statistical fluctuation can noticeably swing its content away from this average. Though statistical fluctuation happens in any system size, large fluctuations (as scaled by the size of the system) are more likely in small size systems. Accordingly, one child model is designed to be small (with about 1 to 20 atoms) on purpose so that it becomes variable to chance. When a fluctuation causes the small child model to contain nearly all "bad" atoms, its sibling (a large child model) will contain slightly more "good" atoms than the parent, and an expansion from this sibling likely leads to a slight improvement on the tentative full model. On the other hand, when a fluctuation causes the small child model to contain nearly all “good” atoms and the parent model is still poorly determined, an expansion from this good small child model may greatly improve the tentative full model. These expectations have been verified by test calculations (see section 6).

With all above considerations in mind, the following scheme of splitting a parent model has been designed. Let $N$ be the number of atoms in the parent model. Calculate $p_1 = 20/N$. If $p_1 > 0.5$, take $p_1 = 0.5$. Generate a random number $p_0$ that is uniformly distributed from 0 to $p_1$. For each atom in the parent model, generate a random number $p$ that is uniformly distributed from 0 to 1. If $p > p_0$, assign this atom to child model 1; otherwise, assign it to child model 2. If either of the child models happens to be empty, try another split.

Figure 1 shows a general flow chart of an sR1 calculation in lottery mode. The whole calculation starts from an initial starting model, which is either a single atom or one/two/a few fragments of a known structure oriented and translationally positioned by the partial-structure R1 (pR1) method (Zhang & Donahue, 2024). This starting model is expanded to model 0 by one cycle of the sR1 calculation. After deleting the likely bad atoms (namely those when introduced caused the approximate R1 to rise) model 0 is split into child models 1 and 2 by the above-mentioned scheme. Expand child model 1 into model 1 by one cycle of the sR1 calculation. If model 1 is an improvement over model 0 (see the next paragraph on how this comparison is made), take model 1 as the new model 0 and move on to the next lottery cycle. Otherwise, expand child model 2 into model 2 by one cycle of the sR1 calculation. If model 2 is an improvement over model 0, take model 2 as the new model 0 and move on to the next lottery cycle. Otherwise, start the next lottery cycle with the unchanged model 0. Note that the lottery cycles go indefinitely. A user checks the improvement of model 0 from time to time and decides when to terminate the calculation.

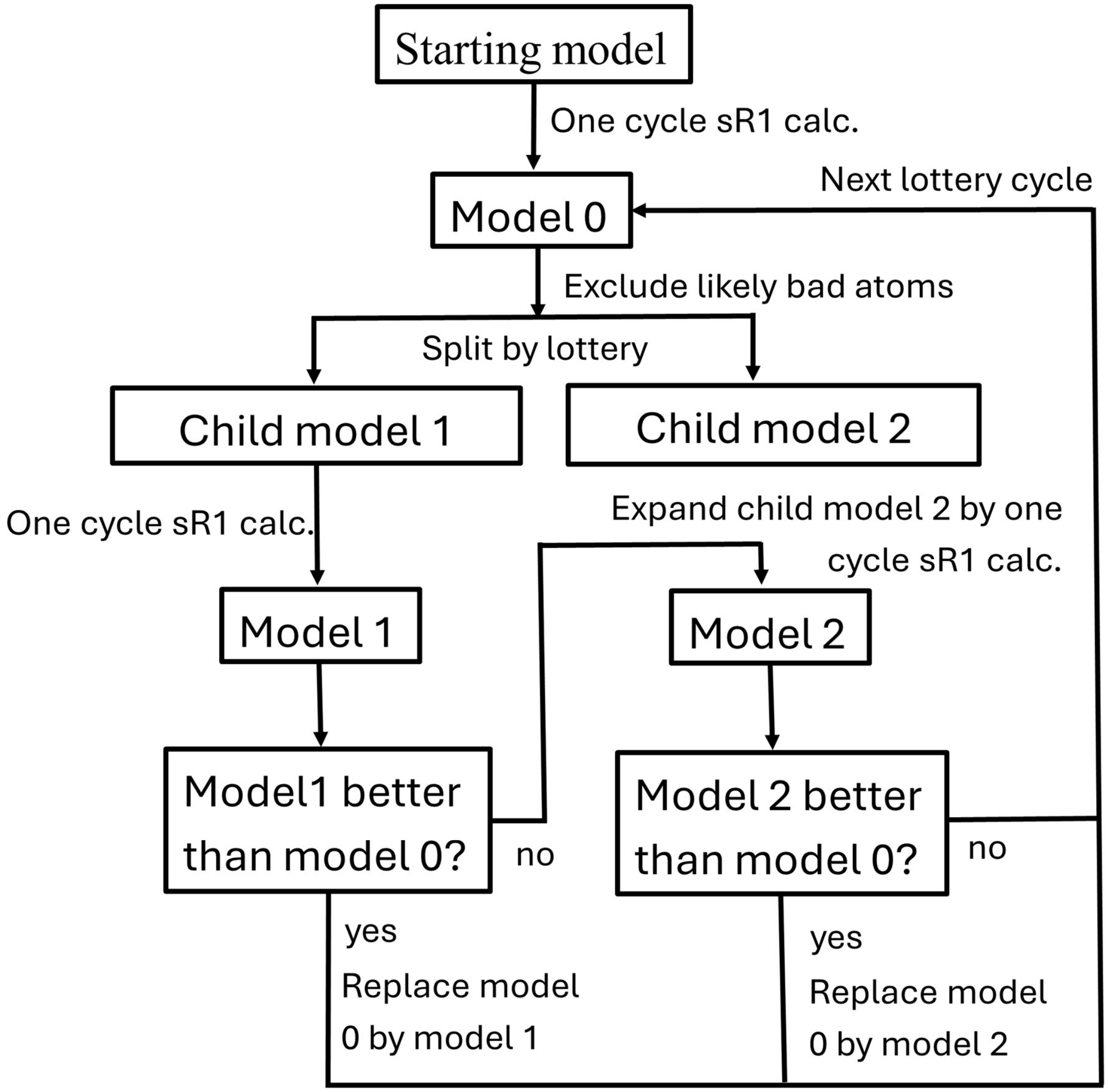

**Figure 1** A general flow chart of an sR1 calculation in a lottery mode.

One thing remains unclear though. How can it be seen whether model 1 (or 2) is an improvement over model 0? When a model grows by adding atoms one by one, the approximate R1 usually drops, but sometimes it increases. By intuition, the minimum approximate R1 ever reached is a good measure of how good the resulting model (after excluding the likely bad atoms as described above) is. This standard has been adopted for comparing models. If the minimum approximate R1 that has been reached when building model 1 is lower than that when building model 0, model 1 is an improvement over model 0. The same rule is applied for comparing model 2 with model 0.

#### 5. Test calculations

The crystallographic information of the samples used in the test calculations is listed in Table 1. The samples will be referred to by their sample numbers.

**Table 1** Crystallographic information of the samples used in the test calculations

| Sample | Formula (excluding H) | Z | non-hydrogen atoms in cell | Data Resolution (Å) | a(Å) | b(Å) | c(Å) | α(°) | β(°) | $\gamma$(°) | Space Group |
|---|---|---|---|---|---|---|---|---|---|---|---|
| **1** | $FN_2C_{13}$ | 8 | 128 | 12.1 – 0.79 | 8.45 | 9.93 | 24.20 | 90.00 | 90.00 | 90.00 | Pbca |
| **2** | $O_2C_{29}$ | 8 | 248 | 10.7 – 1.00 | 44.17 | 10.73 | 8.93 | 90.00 | 90.00 | 90.00 | Pbcn |
| **3** | $F_2O_4N_2C_{52}$ | 4 | 240 | 24.2 – 0.90 | 12.27 | 24.23 | 15.42 | 90.00 | 96.94 | 90.00 | $P2_1/c$ |
| **4** | $PdS_6P_4C_{68}$ | 4 | 316 | 10.5 – 0.85 | 23.32 | 14.45 | 22.39 | 90.00 | 115.22 | 90.00 | $P2_1/c$ |

In the following test calculations, each resulting model was compared to a "correct" model. Here is how such a "correct" model was obtained. A structure was "solved" by the program *SHELXT* (Sheldrick, 2015a). After going through "refinement" with the program *SHELXL* (Sheldrick, 2015b) and deleting H atoms, the structure was expanded to the P1 space group. This expanded structure is the "correct" model. A comparison between a resulting model with a "correct" model helped to verify that the sR1 method yielded approximately the same model as the program *SHELXT* did.

### 5.1. One typical example

Typically, the sR1 method can solve a small molecule structure by starting from a single atom (Zhang & Donahue, 2024). Sample 1 is one such typical example. As seen in Table 1, sample 1 has 128 non-hydrogen atoms in its unit cell. As shown in figure 2, when running in lottery mode, the sR1 calculation can reach a final model of sample 1 by the end of lottery cycle 33. This final model, when compared to the correct model, has all 128 atoms being located within 0.5 Å of their corresponding correct locations.

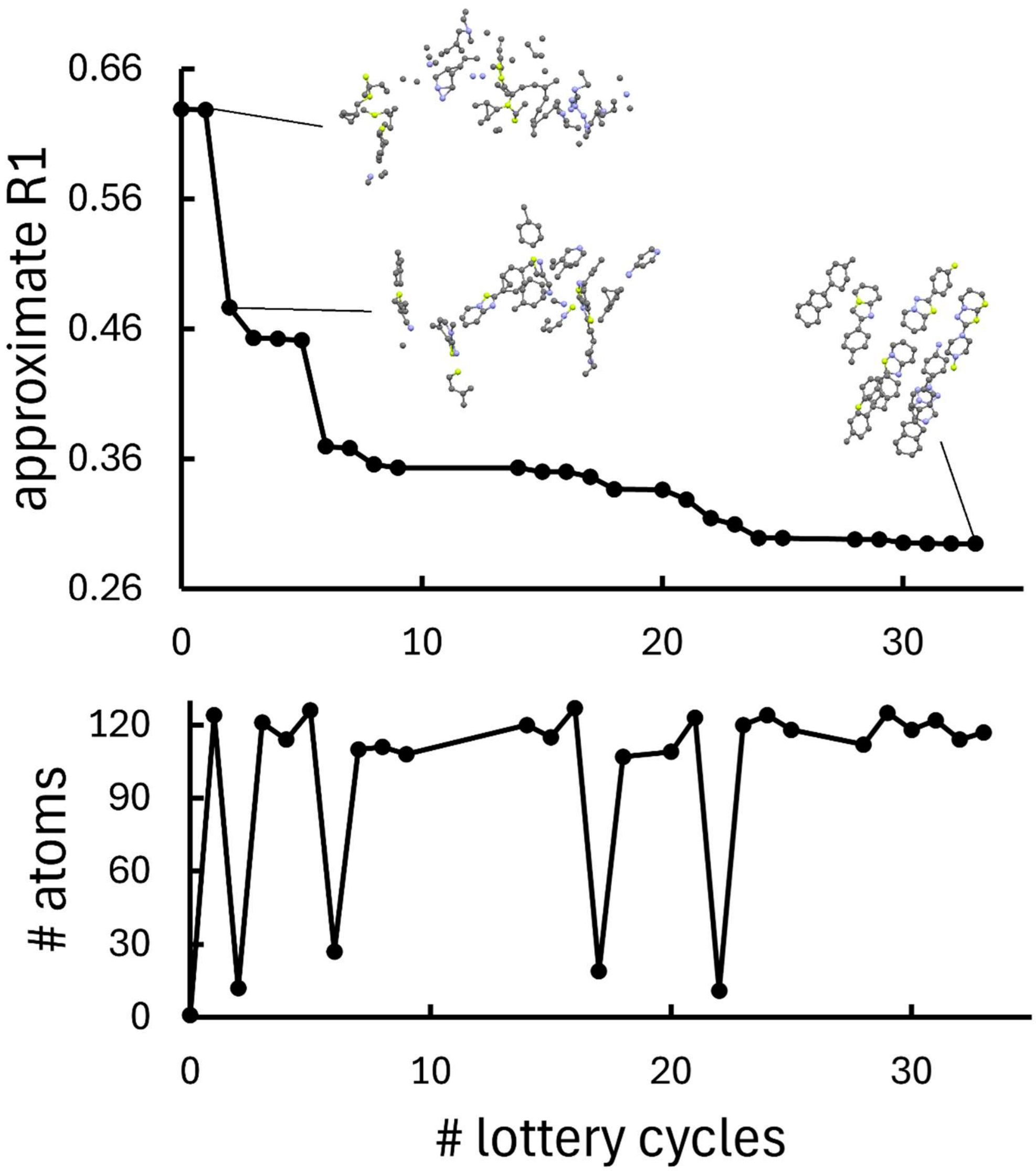

**Figure 2** Solving the structure of sample 1 by running the sR1 calculation in lottery mode, starting from a single F atom at (0.3, 0.3, 0.3). The top chart shows how the minimum approximate R1 that has been reached in each lottery cycle drops when the calculation proceeds. The bottom chart shows the size of the starting partial model of each lottery cycle. The inserts show the model structures at the end of cycles 1, 2, and 33, respectively.

#### 5.2. Another typical example but with more atoms

Sample 2 is another typical example for which the sR1 calculation can start from a single atom. The structure of sample 2 is much larger than that of sample 1. As seen in Table 1, sample 2 has 248 non-hydrogen atoms in its unit cell compared to sample 1's 128. Figure S1 of the supporting information shows that, by the end of lottery cycle 48, the result of solving sample 2 is close to complete.

Compared with the correct model, this result has 239 atoms located within 0.5 Å of their corresponding correct locations, with only 9 atoms being misplaced.

### 5.3. An example of a difficult case

The author has tested the sR1 method over around 224 datasets (see a list within supporting information). For most cases the calculation is straightforward by starting from a single atom at (0.3, 0.3, 0.3). For some of these straightforward cases the structure is revealed by one sR1 cycle, but most of these cases require multiple cycles with a user's guidance of selecting fragments and/or deleting ghost atoms. Then, there are a few difficult cases in which a single sR1 cycle starting from a single atom at (0.3, 0.3, 0.3) does not reveal any recognizable fragments. For these difficult cases the sR1 calculation needs to be jumpstarted by using the partial-structure R1 (pR1) method to orient and translationally position one/two/a few fragments of a known structure (Zhang & Donahue, 2024). Sample 3 is one of these difficult cases. Figure 3 shows the general steps of solving the structure of sample 3. The first step is to use the pR1 method to orient and translationally position two benzene rings. The second step is to perform the lottery cycles by starting from the partial model of the two benzene rings. Figure S2 of the supporting information shows the details of this step. The result of these lottery cycles contains some ghost atoms. However, it should not be expected that the sR1 method can clean out all these ghost atoms as some of these "ghost" atoms are real – they arise from other unknown diffraction centers like disordered solvent, etc. The main goal of solving a structure is to reveal the primary structure. So, these "ghost" atoms are deleted. The third step uses a simple trick to discover the missing atoms of the primary structure. This simple trick can be called the bond length guided sR1 method. Though the author came across this simple idea with no knowledge of the existence of other programs, as one reviewer has pointed out, it should be acknowledged that similar ideas have been well exploited in other programs, including *ARPWarp* (Lamzin *et al.*, 2012), *Coot* (Emsley *et al.*, 2010) and *PHENIX AutoBuild* wizard (Terwilliger *et al.*, 2008). The underlying principle of this idea is to let A be an atom that has already been located, and Q be an undetermined atom which is expected to bond to atom A with an approximate bond length $r$. The approximate C-C bond length 1.39 Å was used in this work. A grid has been set up within the unit cell with step size 0.4 Å. To coarsely locate atom Q, only the grid points that are located within a spherical shell, which centers at atom A and has radii $r - \Delta r \sim r + \Delta r$, need to be tested. Here, $\Delta r$ takes the value of either 0.3, or 0.4, or 0.5 Å (in this report, 0.3 Å was used). Other than limiting the initial search range for positioning atom Q, the rest of the bond length guided sR1 calculation is the same as the normal sR1 procedure (Zhang & Donahue, 2024). The third step of figure 3 shows the final model of sample 3 after discovering the missing atoms by the bond length guided sR1 method. Compared with the correct model, this result has all 240 atoms being located within 0.5 Å of their corresponding correct locations, with no atoms being misplaced.

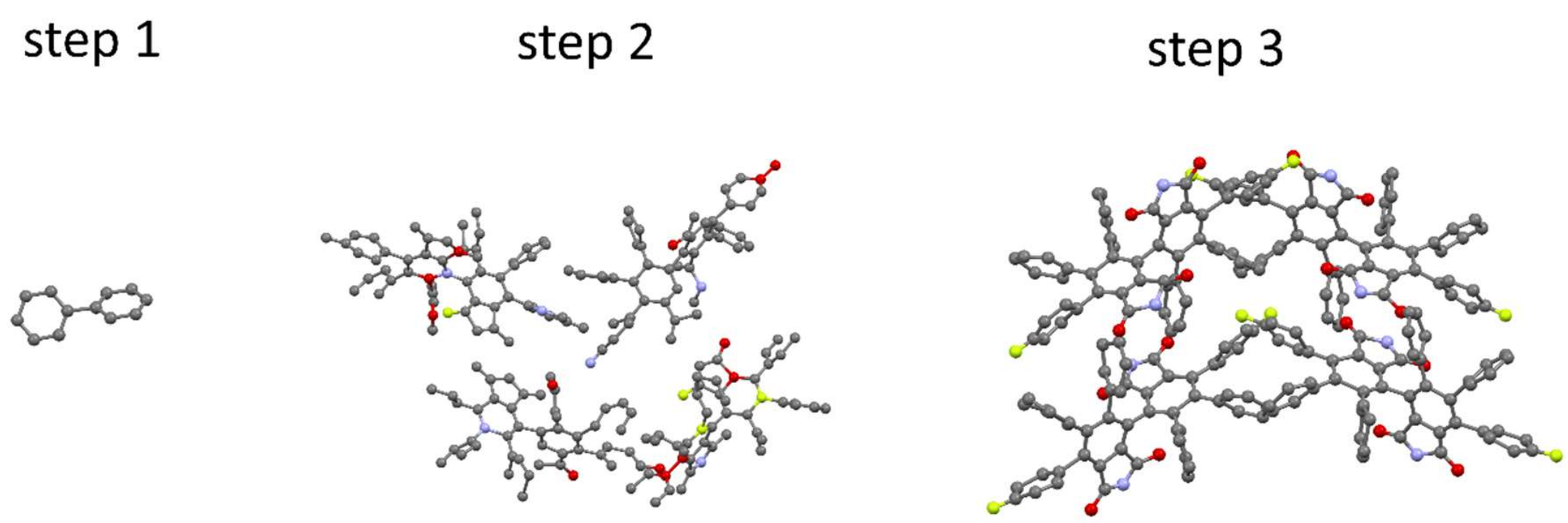

**Figure 3** General steps of solving the structure of sample 3. Step 1: the partial model formed by using the pR1 method to orient and translationally position two benzene rings. Step 2: the resulting partial structure after running 36 lottery cycles, starting from the partial structure of the two benzene rings. Note that the ghost atoms are deleted from this resulting partial structure. Step 3: the final model after using the bond length guided sR1 method to add the missing atoms. Note that the atom types have been manually corrected.

#### 5.4. A structure containing many heavy atoms

Sample 4 is an example of a structure containing many heavy atoms. The unit cell of this structure contains 4 Pd atoms, 24 S atoms, 16 P atoms, and 272 C atoms (see table 1). There are 44 heavy atoms in this structure. For such a structure, it makes sense to first determine the heavy atom substructure and then add the light atoms later. Figure 4 shows the general steps of solving the structure of sample 4. The first step is to solve the heavy atom substructure by running the sR1 calculation in lottery mode, starting with a single S atom at (0.3, 0.3, 0.3). The details of the lottery cycles are shown in figure S3 in the supporting information. The second step is to add the C atoms by one cycle of the sR1 calculation, starting with the heavy atom substructure that has been discovered in the first step. Step 2 results in some ghost C atoms, which are deleted. In step 3, the missing C atoms are discovered by the bond length guided sR1 method. The final model, when compared to the correct model, has 315 atoms being located within 0.5 Å of their corresponding correct locations, with only one C atom being misplaced.

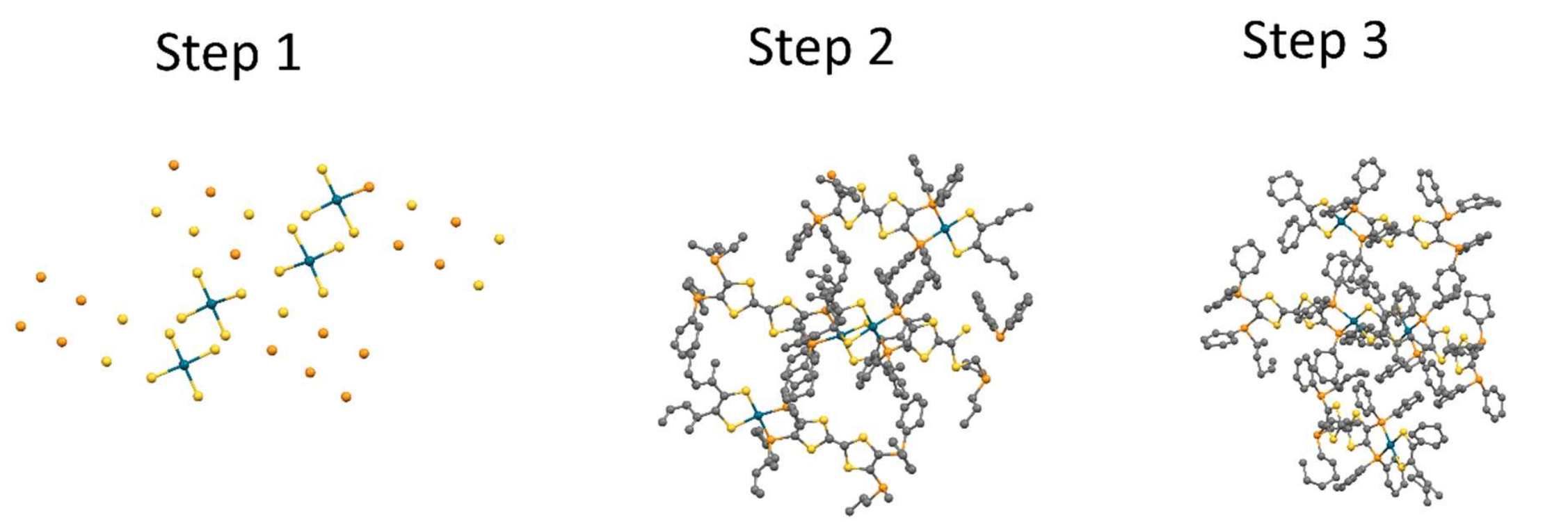

**Figure 4** General steps of solving the structure of sample 4. Step 1: the heavy atom substructure discovered by running 6 lottery cycles of the sR1 calculation. Step 2: the resulting partial structure after running one cycle sR1 calculation, starting from the heavy atom substructure. Note that ghost C atoms are deleted from this partial structure. Step 3: the final model after using the bond length guided sR1 method to add the missing C atoms.

### 6. Discussions and conclusions

#### 6.1. Success rate of the lottery cycles

Figure 2 shows that for solving the structure of sample 1, a total of 33 lottery cycles have been tried. Among those 33 cycles, 26 led to a drop of the approximate R1 (shown on the charts), while 7 were skipped because of failing to improve the model. So, the success rate of the lottery cycles in this case was 26 over 33. Similarly, the success rate for sample 2 was 28 over 48 (Figure S1), for sample 3 25 over 36 (Figure S2), and for sample 4 6 over 6 (Figure S3). The success rates were higher than 50%. One reason of these high success rates was that in each lottery cycle there were two child models to try. When one child is “worse” than the parent, the other must be “better” than the parent, and its expansion likely leads to a success. Another likely reason is that one child model has a small size of 1 to 20 atoms. For a small child model, the chance of large statistical fluctuation is high, making the model either contain nearly all “bad” atoms or nearly all “good” atoms. When the small child model contains nearly all “bad” atoms, its large sibling may lead to a small improvement on the solution. On the other hand, when the small child model contains nearly all “good” atoms, it itself can lead to an improvement. The improvement is either great if the parent model is still poorly determined or slight if the parent model has been well determined (see the next sub-section).

#### 6.2. The contribution of small child models vs that of large child models

As seen in the bottom chart of figure 2, for sample 1, out of the 26 successful lottery cycles, 22 were started with large child models. Only 4 were started with small child models. However, the small child models provided most of the improvement on the solution, namely providing the most drop in the approximate R1 (see top chart of figure 2). Among the four occasions of small child models at cycles 2, 6, 18, and 22, cycles 2 and 6 happened when the parent model was still poorly determined. This subsequently caused the approximate R1 to drop dramatically (see top chart of figure 2). On the other hand, cycles 18 and 22 happened when the parent model was well determined, so only small improvements were made at these cycles. As the inserts in figure 2 indicate, the resulting model at the end of cycle 1 had very few recognizable fragments; however, by the end of cycle 2, many recognizable fragments suddenly appeared. This observation of the small child model initiating successful cycles happening less frequently but contributing greatly on improving the solution is also observed for samples 2 (figure S1) and 4 (figure S3). For sample 3 (figure S2), all 25 successful cycles were initiated by large child models. The reason is simple for this occurrence. The calculation of sample 3 was started with a partial model of two benzene rings that were oriented and translationally positioned by the pR1 method. Therefore, it needed to be extremely lucky for a small child model to beat such a starting partial model where a new rebuild from scratch could be accepted.

A cycle started with a large child model aims at making a small improvement, while a cycle started with a small child model aims at rebuilding the model from scratch. The above observations indicate that when the model is still poorly determined, a rebuild can dramatically improve it. For this reason, small child models are important. Though they less frequently lead to improvement, when they do the effects sometimes can be revolutionary.

### 6.3. The need of a fast-computing tool

For the difficult case of sample 3, is it possible to start the sR1 calculation with a partial model of a single atom instead of using the pR1 method to put two benzene rings? The answer to this question is "Yes". Figure S4 in the supporting information shows one successful trial of solving the structure of sample 3 by running the sR1 method in the lottery mode, starting with a single F atom at (0.3, 0.3, 0.3). Figure S4 should be compared with figure S2. It is seen that starting from single F atom needed 243 lottery cycles, while starting from two benzene rings needed only 36 lottery cycles. The comparison of the actual computing time (running on a Microsoft Surface Pro 7, Intel Core i7-065G7 CPU @ 1.30 GHz, 16.0 GB RAM) was about 75 hours for the 243 cycles vs about 7 hours for the 36 lottery cycles. Obviously, neither calculation is practical to perform, especially the one starting with a single atom. However, the sR1 method is well suited for parallel calculation. The values of the approximate R1 at all grid points can be calculated simultaneously. So, the whole sR1 map can be analyzed within a split of a millisecond, and the whole calculation can go extremely fast if sufficient computing power is available. If such fast-computing tools are used, all these calculations can be comfortably performed.

### 6.4. A universal way of starting the sR1 calculation of a "difficult" case?

There is one more thing about figure S4 in the supporting information that is worth attention. The inserts show that at the end of cycle 210 the resulting model still had no recognizable fragments, but by the end of cycle 214 many recognizable fragments suddenly appeared. Note that cycles 211, 212 and 213 were skipped because of failing to improve the model. More shockingly, the starting partial model of cycle 214 was a single N atom at (0.8333, 0.2333, 0.1184). By some magic, this location made this sR1 cycle very successful. In comparison, the initial starting model of a single F atom at (0.3, 0.3, 0.3) was unsuccessful. This peculiar phenomenon suggests that there is subtlety on where to put the initial single atom. By the experience of applying the sR1 method to about 224 datasets, the default initial position (0.3, 0.3, 0.3) works for most cases. The observation that (0.8333, 0.2333, 0.1184) works much better than the default (0.3, 0.3, 0.3) for sample 3 indicates that, the reason that sample 3 was regarded as a "difficult" case might simply be because the default choice happened to be a bad choice for this dataset. Why does where to put the starting single atom matter? A grid of 0.4 Å step size is set up within the unit cell. The grid points are fixed within the unit cell. However, the location of the sR1 holes depends on the location of the single atom. Therefore, how well the grid points can coarsely catch the sR1 holes depends on the location of the single atom. There must be an optimal location of this single atom such that one cycle of the sR1 calculation can lead to the best tentative full model, as measured by the lowest of the minimum approximate R1. Note that each trial reaches a minimum approximate R1, and the trial with the lowest resulting minimum approximate R1 is the best trial. There is no need to search for this optimal location in the whole unit cell. Instead, a search within a small box of 0.4 Å by 0.4 Å by 0.4 Å with the corner of the lowest coordinates at (0.3, 0.3, 0.3) is sufficient. A coarse search with 4 by 4 by 4 (= 64) local grid points within this box for sample 3 yielded an optimal location of (0.325, 0.3042, 0.3). The structure of sample 3 can be successfully solved by starting lottery cycles with a partial model of a single F atom at (0.325, 0.3042, 0.3). It was observed that a dramatic drop of the approximate R1 happened on cycle 26, a cycle started with a small child model of 20 atoms. A good final model was reached at the end of cycle 85 (see more details in figure S5 of the supporting information). So, for a "difficult" case like the case of sample 3, a universal way to start the sR1 calculation might be to find an optimal initial location for the starting single atom. This approach seems general. It might work even when no fragments of known structure are available for running the pR1 method to jumpstart the sR1 calculation. However, this idea currently is only at the initial stage, and further tests are needed.

### 6.5. The effect of data resolution

The data resolution limits of samples 1 to 4 are given in Table 1. Of the four samples, 1 and 4 have higher resolutions (0.79 Å and 0.85 Å, respectively) while 2 and 3 have lower resolutions (1.0 Å and 0.90 Å, respectively). Interestingly, though sample 3 has slightly higher resolution than sample 2, 3 was regarded as a "difficult" case while 2 was not.

The effect of data resolution on the effectiveness of the sR1 method has been studied in the previous paper (Zhang & Donahue, 2024). Essentially, lower data resolution makes a structure harder to solve. When data resolution is high, a structure can be straightforwardly solved by starting from a single atom at (0.3, 0.3, 0.3). However, when data resolution is low, the case becomes more difficult. Before the lottery mode was designed, a difficult case required either intelligently guessing a correct fragment or orienting and positioning fragments of a known structure by the pR1 method to jumpstart the sR1 calculation. Now, with the lottery mode, even difficult cases can be solved by starting from a single atom (see section 6.3). Take sample 3 in the previous paper (Zhang & Donahue, 2024) as an example. Previously, when data resolution was truncated to 1 Å, different avenues had to be taken to solve the structure. For instance, one route required a fragment of 4 C atoms to be guessed. In another route, a benzene ring had to be correctly oriented through the pR1 method. Now, using the lottery mode, as shown in Figure S6 in the supporting information, this difficult case (when data resolution is truncated to 1 Å) can be directly solved by starting from a single C atom at (0.3, 0.3, 0.3). However, the structure was solved only after 113 lottery cycles in addition to the initial sR1 cycle. In comparison, if the data resolution is 0.77 Å (Zhang & Donahue, 2024) the structure can be solved by the initial sR1 cycle only. So, clearly low data resolution makes a structure much harder to solve.

### 6.6. How well does the lottery mode work for various datasets?

The author has tested the sR1 method on about 224 datasets. A list is given in the supporting information. After solving a structure with the sR1 method, the *Platon* program (Spek, 2020) was used to determine the space group and to help to find the asymmetric unit (see an example in the supporting information). Finally, the structure was refined by the *SHELXL* program (Sheldrick, 2015b). The final R-values for reflections of Fo > 4 sig(Fo) are listed in the same table in the supporting information. The main factor affecting an R-value is the data quality of a sample. As can be seen in the list, most datasets (184 of them) have an R-value smaller than 0.1. There are 22 datasets that have an R-value between 0.1 and 0.16. These datasets are of intermediate quality. Then there are 18 datasets that have an R-value higher than 0.16. These are low-quality datasets. The sR1 method was less efficient for solving these datasets. Nevertheless, the method extracted useful information of the underlying structures from these datasets.

About 85 datasets have been tested with the lottery mode (as labeled by letter “L” in the list). For most cases out of the 224 datasets the first sR1 cycle revealed sizable recognizable partial structure. For these cases, the lottery mode was expected to work smoothly. Indeed, tests have confirmed this intuition. Only when the initial sR1 cycle shows sign of trouble (namely when there are nearly no recognizable fragments) does the lottery mode have great value. In the 224 datasets, there were a few such cases where the lottery mode necessitated use, and the results were all successful. The selected samples 1, 2 and 3 were among the most difficult ones in this group. Sample 4 was selected as a representative structure for those with many heavy atoms. Usually, structures containing many heavy atoms like sample 4 are readily solved by the sR1 method. In fact, they also fall in the group in which the first sR1 cycle reveals sizable recognizable partial structure, therefore, no trouble is expected for these cases when applying the lottery mode. For a few datasets that could easily be solved by the sR1 method in the "intelligent" mode and could also be trivially done through the lottery mode, these datasets were still tested using the lottery mode out of curiosity. However, these cases that were subjected to the lottery mode were intentionally made more difficult. For example, instead of starting from a single atom, the model started with all atoms being randomly positioned in the unit cell. Despite the increased difficulty, the lottery mode still worked well in these tests. However, this outcome should have been "expected" because the lottery mode has been designed to have a small child model for rebuilding a model from scratch.

This new method of running sR1 in lottery mode indeed works "by chance". Tests have shown that if the same calculation is repeated a few times, starting from the same partial model and ending at approximately the same final model, the method takes a different path each time. Some paths need more lottery cycles than others. Though usually the variation is small, sometime the variation can be as wild as "more than 100 cycles vs about 30 cycles". Talking about the general test calculations, for most cases a calculation finishes within a hundred lottery cycles. But in some cases, it needs to run a few hundred cycles. Sometime a sudden improvement of the model happens near the end. But sometime the calculation only gradually approaches the final solution. If one notices that the primary structure is obviously missing a few atoms but the method "refuses" to add atoms there, it is better to stop the cycles and use the bond length guided sR1 method to add the missing atoms instead. The cause of this is that some "ghost" sites are competing with the primary structure.

It should be acknowledged that when there are recognizable fragments the "intelligent" mode is the most efficient way to guide the calculation towards the final solution. In future when "intelligent" pattern recognition is automated, the intelligent mode and the lottery mode will combine. When there are no recognizable fragments the lottery mode is used. Once there are recognizable fragments, the calculation switches to intelligent mode. However, after the structure is complete, running extra lottery cycles can improve the model further. For example, distorted benzene rings can be improved, etc. For this purpose, the size of the small child model should be increased so that the calculation is more efficient. Another possibility, as one reviewer has pointed out, is to use the sR1 method for phasing purpose: the sR1 method only needs to determine a partial structure to provide an initial phasing and the usual dual space recycling process can expand the partial structure into a final full model.

### 6.7. Conclusions

Based on the above observations, the main conclusions of this study can be summarized as follows: (1) The designed lottery mode can drive the sR1 calculation toward correct structure solutions. (2) This new method, namely, carrying out the sR1 calculation in a lottery mode, is useful for a care-free application of the sR1 method. (3) A fast-computing tool is needed for further exploration of the sR1 method. Slow rate of calculation with Surface Pro units is limiting the author's ability of testing various new ideas of developing the sR1 method. However, the method itself can go extremely fast if parallel programming is fully exploited and a dedicating supercomputer is accessible.

**Acknowledgements** Professor James P. Donahue, Professor Robert A. Pascal and Professor Joel T. Mague provided the crystal samples used in this work.

**Availability of the source code:** see section S10 of the supporting information.

## References

Bott, R. & Sarma, R. (1976). *J. Mol. Biol.* **106**, 1037-1046.

Bricogne, G. (1992). *Proceedings of the CCP4 Study Weekend. Molecular Replacement*, edited by W. Wolf, E. J. Dodson & S. Gover, pp. 62-75. Warrington: Daresbury Laboratory.

Burla, M. C., Carrozzini, B., Cascarano, G. L., Giacovazzo, C. & Polidori, G. (2018). *Acta Cryst.* A**74,** 123-130.

Burla, M. C., Carrozzini, B., Cascarano, G. L., Giacovazzo, C. & Polidori, G. (2020). *Acta Cryst.* D**76**, 9-18.

Caliandro, R., Carrozzini, B., Cascarano, G. L., Giacovazzo, C., Mazzone, A. & Siliqi, D. (2009). *Acta Cryst.* A**65**, 512-527.

Crowther, R. A. (1972). In *The Molecular Replacement Method*, Ed. M. G. Rossmann, Gorden & Breach, New York, pp. 173-178.

DeTitta, G. T., Weeks, C. M., Thuman, P., Miller, R. & Hauptman, H. A. (1994). *Acta Cryst.* A**50**, 203-210.

Emsley, P., Lohkamp, B., Scott, W. G. & Cowtan, K. (2010). *Acta Cryst.* D**66**, 486–501.

Eventoff, W., Hackert, M. L. & Rossmann, M. G. (1975). *J. Mol. Biol.* **98**, 249-258.

Giacovazzo, C. (2014). *Phasing in Crystallography – A Modern Perspective.* Oxford: Oxford University Press.

Gorelik, T. E., Lukat, P., Kleeberg, C., Blankenfeldt, W. & Mueller, R. (2023). *Acta Cryst*. A**79**, 504-514.

Kinneging, A. J. & de Graaff, R. A. G. (1984). *J. Appl. Cryst.* **17**, 364-366.

Lamzin, V. S., Perrakis, A. & Wilson, K. S. (2012). *International Tables for Crystallography. Vol. F, Crystallography of Biological Macromolecules*, 2nd online ed., edited by E. Arnold, D. M. Himmel and M. G. Rossmann, pp. 525–528. Dordrecht: Kluwer Academic Publishers. McCoy, A. J. (2004). *Acta Cryst.* D**60**, 2169-2183.

McCoy, A. J. (2017). In *Protein Crystallography: Methods and Protocols*, Ed. A. Wlodawer, Z. Dauter & M. Jaskolski, Humana Press, New York, pp. 421-454.

McCoy, A. J., Grosse-Kunstleve, R. W., Adams, P. D., Winn, M. D., Storoni, L. C. & Read, R. J. (2007). *J. Appl. Cryst*. **40**, 658–674.

McCoy, A.J., Oeffner, R. D., Wrobel, A. G., Ojala, J. R., Tryggvason, K., Lohkamp, B. & Read, R. J. (2017). Ab initio solution of macromolecular crystal structures without direct methods. *Proc. Natl. Acad. Sci. USA* **114**, 3637-3641.

Miller, R., DeTitta, G. T., Jones, R., Langs, D. A., Weeks, C. M. & Hauptman, H. A. (1993). *Science*, **259**, 1430-1433.

Read, R. J. (2001). *Acta Cryst.* D**57**, 1373-1382.

Read, R. J. & McCoy, A. J. (2016). *Acta Cryst.* D**72**, 375-387.

Rossmann, M. G. (1972). Editor. *The Molecular Replacement Method*. New York: Gordon & Breach.

Rossmann, M. G. (1990). *Acta Cryst.* A**46**, 73-82.

Rossmann, M. G. & Blow, D. M. (1962). *Acta Cryst*. **15**, 24-31.

Rossmann, M. G., Blow, D. M., Harding, M. M., & Coller, E. (1964). *Acta Cryst.* **17**, 338-342.

Sheldrick, G. M. (2015a). *Acta Cryst.* A**71**, 3-8.

Sheldrick, G. M. (2015b). *Acta Cryst.* C**71**, 3-8.

Sheldrick, G. M. & Gould, R. O. (1995). *Acta Cryst*. B**51**, 423-431.

Spek, A.L. (2020). *Acta Cryst.* E**76**, 1-11.

Terwilliger, T. C., Grosse-Kunstleve, R. W., Atonine, P. V., Moriarty, N. W., Zwart, P. H., Hung, L., Read, R. J. & Adams, P. D. (2008). *Acta Cryst.* D**64**, 61-69.

Weeks, C. M., DeTitta, G. T., Miller, R. & Hauptman, H. A. (1993). *Acta Cryst.* D**49**, 179-181.

Zhang, X. & Donahue, J. P. (2024). *Acta Cryst.* A**80**, 237-248.

# Supporting information

### S1. The algorithm of comparing two models

This algorithm only compares the positions of the atoms in the two models; it disregards the types of atoms. Let $a$, $b$, and $c$ be the length of the unit cell edges. One difficulty of making this comparison is that the two models may have different locations relative to the unit cell edges. The algorithm uses an approximate method to overcome this difficulty: shift all atoms of one model (model A) such that one of its atoms overlaps one atom of the other model (model B). Try this for all atoms in both models. So, the calculation needs to be repeated for $N_A \times N_B$ times, where $N_A$ and $N_B$ are the number of atoms in the models A and B, respectively. Further these calculations also need to be repeated after inverting model B, in case model A matches the inverted version of B better. After repeating the calculation, use the result of the best match that has been discovered.

After overlapping one atom of model A with a particular atom of model B by properly shifting all atoms of model A, we make sure all atoms of both models are located within the unit cell (by adding 1 to or subtracting 1 from the fractional coordinates $x, y, z$ of an atom). With such preparation we are ready to count how many atoms of A are each within $s = 0.5$ Å of an atom in B. To make such comparison quick, we convert the fractional coordinates $x, y, z$ of an atom to integers in the following way:

$$N_x = int(a/s), N_y = int(b/s), N_z = int(c/s)$$

$$I_x = int(xN_x), I_y = int(yN_y), I_z = int(zN_z)$$

Thus, $(I_x, I_y, I_z)$ is the integer coordinate of an atom.

Prepare a mask $M_A(i,j,k)$ for model A, where $i = 0$ to $N_x - 1$, $j = 0$ to $N_y - 1$, and $k = 0$ to $N_z - 1$. $M_A(i,j,k)$ takes value 0 unless $(i,j,k)$ is the integer coordinate of an atom. In that case, it takes value 1. Similarly, $M_B(i,j,k)$ is a mask for model B.

The number of atoms of model A which overlap within 0.5 Å with an atom of B can be calculated as $\sum_{i,j,k} M_A(i,j,k) M_B(i,j,k)$.

**S2. The lottery cycles of solving sample 2**

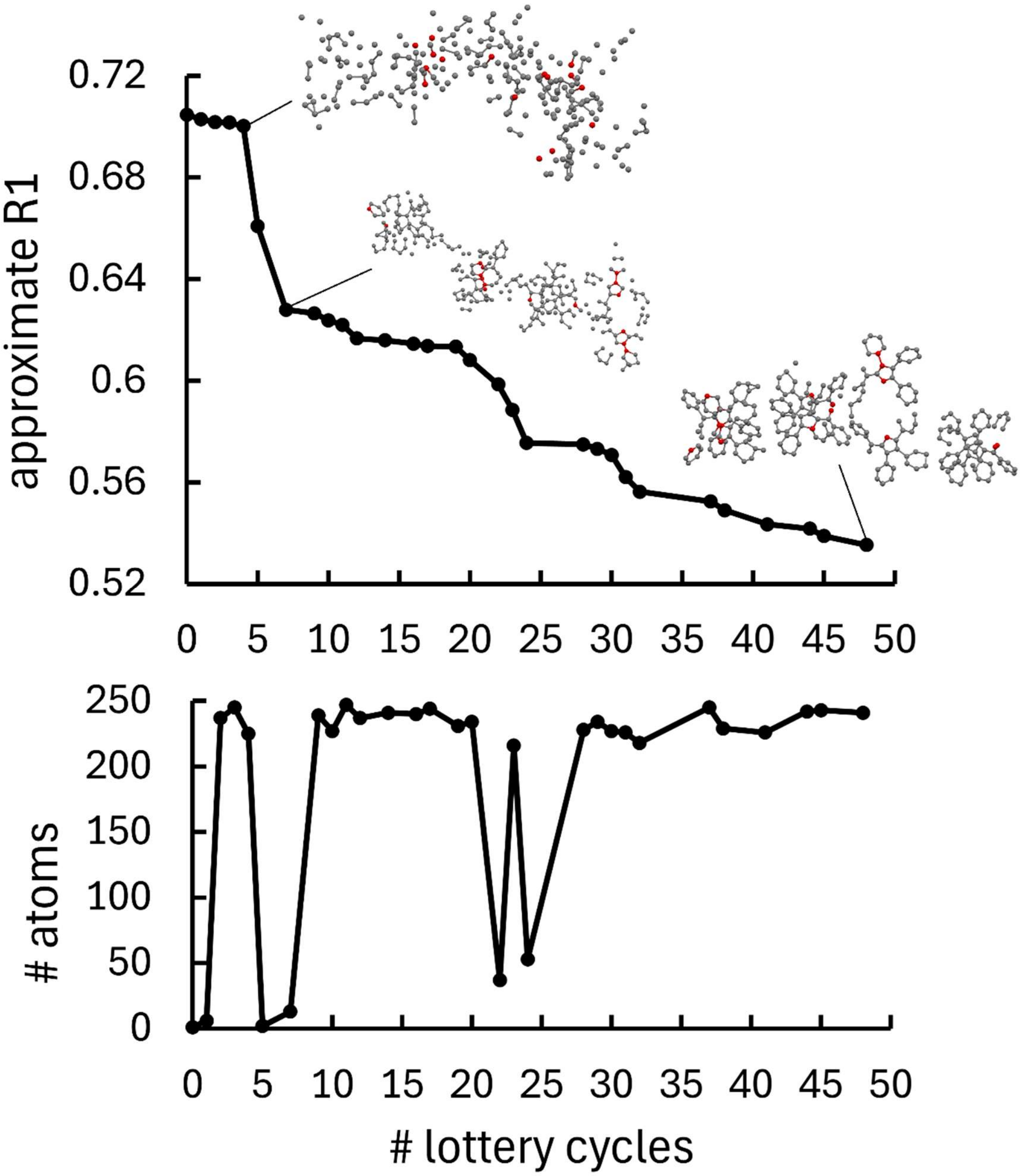

**Figure S1** Solving the structure of sample 2 by running the sR1 calculation in lottery mode, starting from a single O atom at (0.3, 0.3, 0.3). The top chart shows how the minimum approximate R1 that has been reached in each lottery cycle drops when the calculation proceeds. The bottom chart shows the size of the starting partial model of each lottery cycle. The inserts show the model structures at the end of cycles 4, 7, and 48, respectively.

**S3. The lottery cycles of solving sample 3**

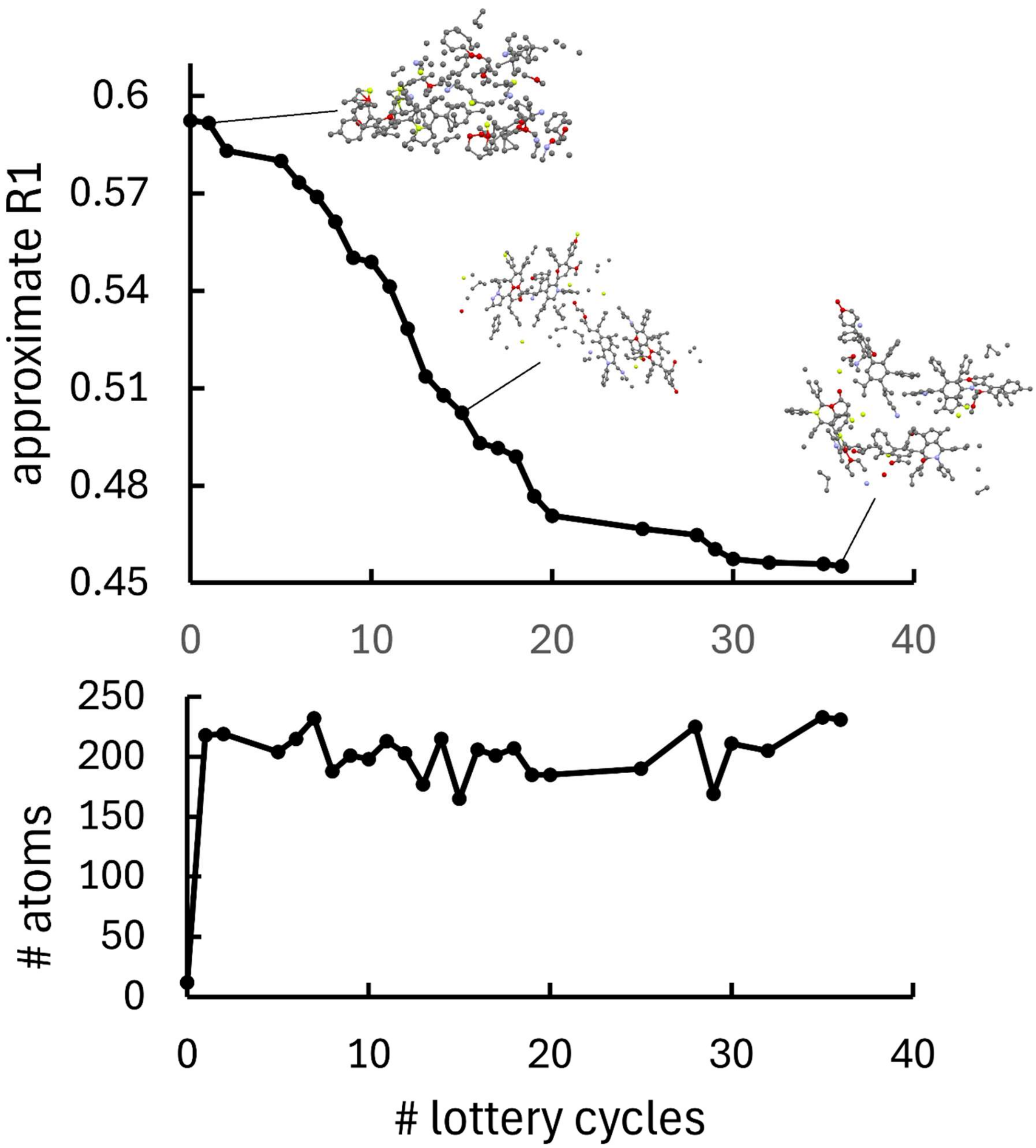

**Figure S2** Details of the lottery cycles for solving the structure of sample 3, starting from the partial structure of two benzene rings. The top chart shows how the minimum approximate R1 reached in each lottery cycle drops when the calculation proceeds. The bottom chart shows the size of the starting partial model of each lottery cycle. The inserts show the model structures at the end of cycles 1, 15 and 36, respectively.

**S4. The lottery cycles of solving sample 4**

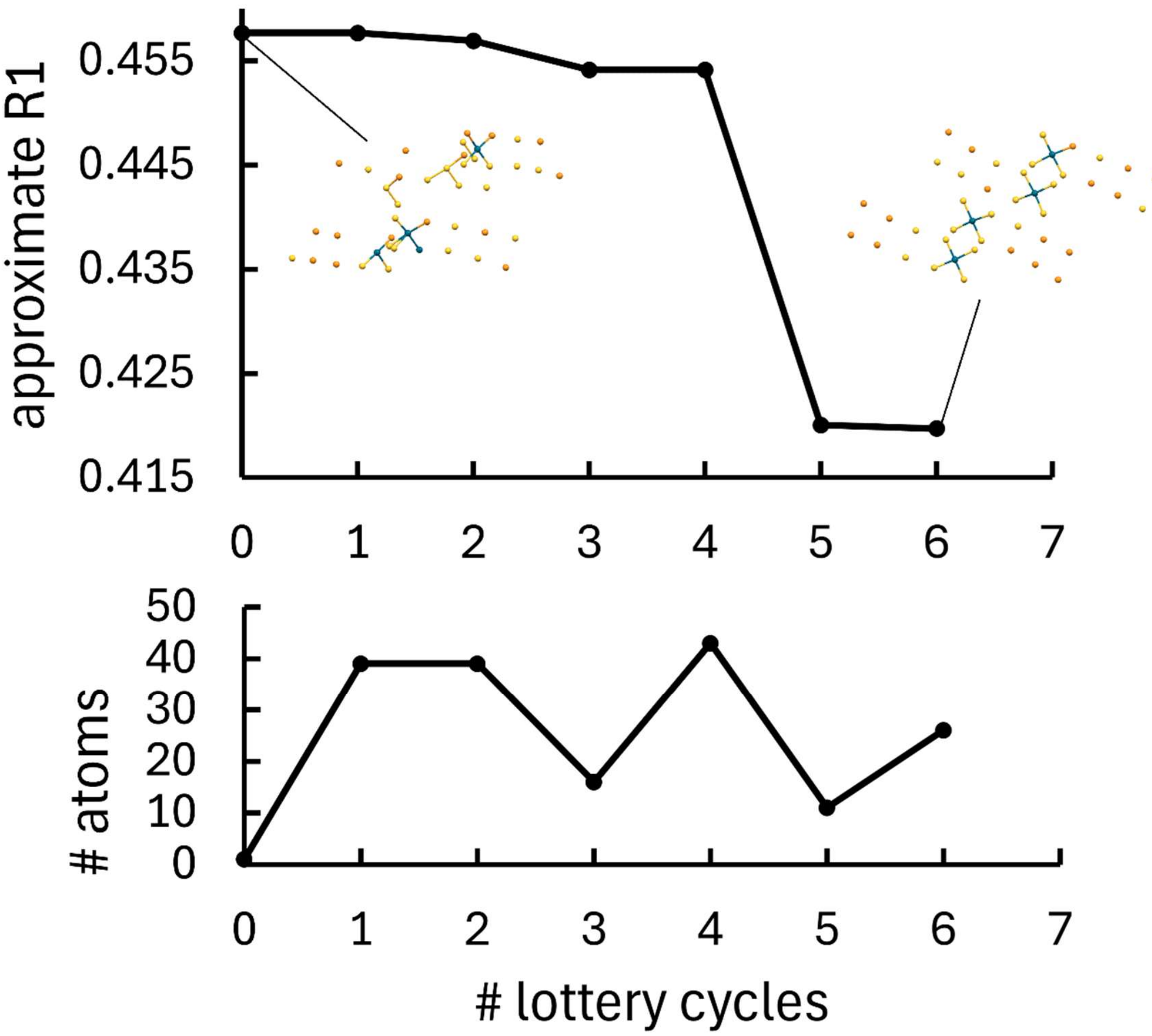

**Figure S3** Details of the lottery cycles for determining the heavy atom substructure of sample 4, starting from a single S atom at (0.3, 0.3, 0.3). The top chart shows how the minimum approximate R1 reached in each lottery cycle drops when the calculation proceeds. The bottom chart shows the size of the starting partial model of each lottery cycle. The inserts show the models of the heavy atom substructure at the end of cycles 0 and 6, respectively.

**S5. The lottery cycles of solving sample 3 by starting from single F atom**

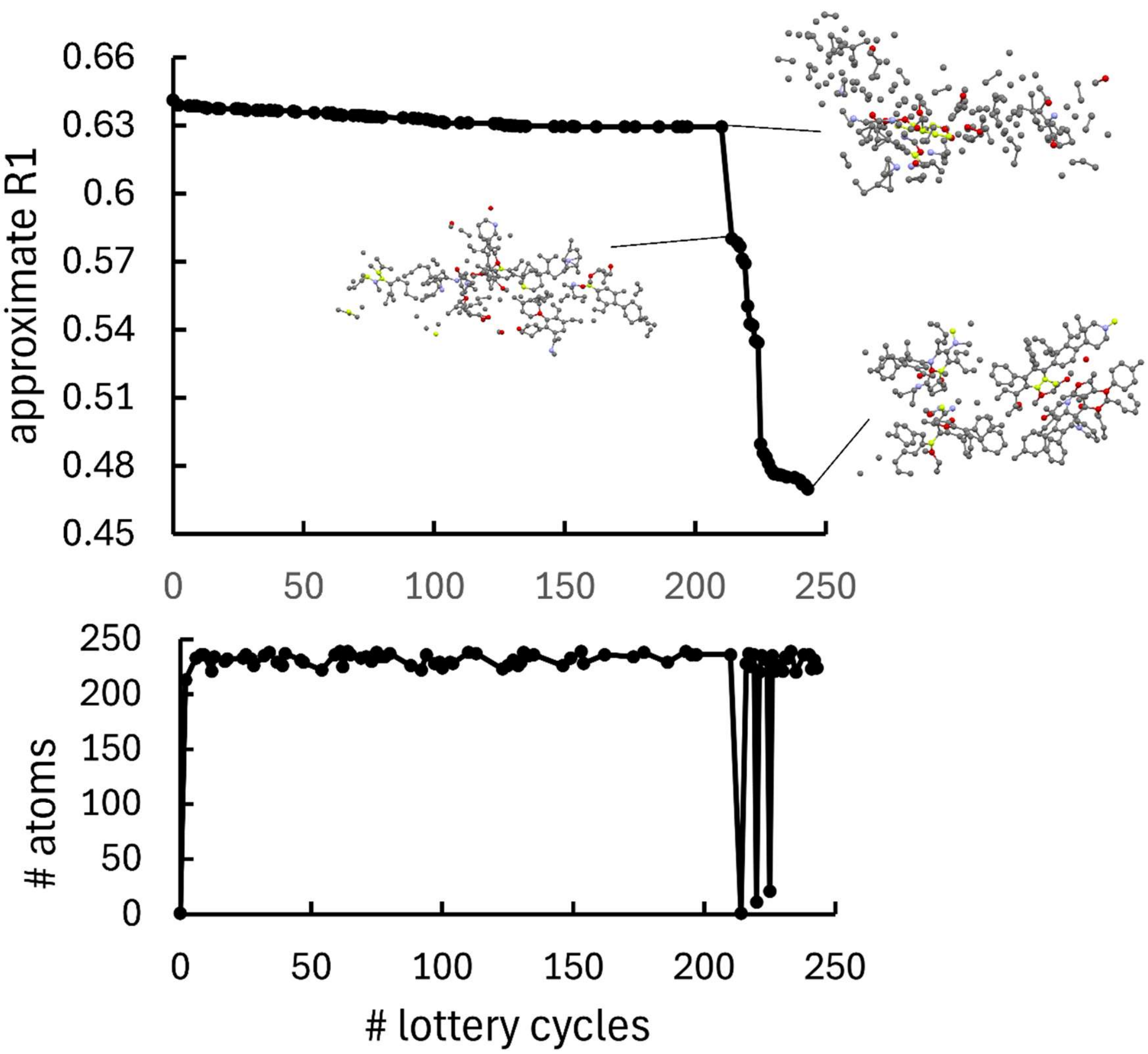

**Figure S4** Solving the structure of sample 3 by running the sR1 calculation in lottery mode, starting from a single F atom at (0.3, 0.3, 0.3). The top chart shows how the minimum approximate R1 in each lottery cycle drops when the calculation proceeds. The bottom chart shows the size of the starting partial model of each lottery cycle. The inserts show the model structures at the end of cycles 210, 214, and 243, respectively.

**S6. A universal way of starting the sR1 calculation of a "difficult" case?**

There is one more thing about figure S4 that is worth attention. The inserts of figure S4 show that at the end of cycle 210 the resulting model still had no recognizable fragments, but by the end of cycle 214 many recognizable fragments suddenly appeared. Note that cycles 211, 212 and 213 were skipped because of failing to improve the model. More shockingly, the starting partial model of cycle 214 was a single N atom at (0.8333, 0.2333, 0.1184). By some magic, this location made this sR1 cycle very successful. In comparison, the initial starting model of a single F atom at (0.3, 0.3, 0.3) was unsuccessful. This peculiar phenomenon suggests that there is subtlety on where to put the initial single atom. By the experience of applying the sR1 method to about 224 datasets, the default initial position (0.3, 0.3, 0.3) works for most cases. The observation that (0.8333, 0.2333, 0.1184) works much better than the default (0.3, 0.3, 0.3) for sample 3 indicates that, the reason that sample 3 was regarded as a “difficult” case might simply was because the default choice happened to be a bad choice for this dataset. Why does where to put the starting single atom matter? A grid of 0.4 Å step size is set up within the unit cell. The grid points are fixed within the unit cell. However, the location of the sR1 holes depends on the location of the single atom. Therefore, how well the grid points can coarsely catch the sR1 holes depends on the location of the single atom. There must be an optimal location of this single atom such that one cycle of the sR1 calculation can lead to the best tentative full model, as measured by the lowest of the minimum approximate R1. Note that each trial reaches a minimum approximate R1, and the trial with the lowest resulting minimum approximate R1 is the best trial. There is no need to search for this optimal location in the whole unit cell. Instead, a search within a small box of 0.4 Å by 0.4 Å by 0.4 Å with the corner of the lowest coordinates at (0.3, 0.3, 0.3) is sufficient. A coarse search with 4 by 4 by 4 (= 64) local grid points within this box for sample 3 yielded an optimal location of (0.325, 0.3042, 0.3). The structure of sample 3 can be successfully solved by starting lottery cycles with a partial model of a single F atom at (0.325, 0.3042, 0.3) (see figure S5).

As shown in figure S5, a dramatic drop of the approximate R1 happened at cycle 26, a cycle started with a small child model of 20 atoms. A good final model was reached at the end of cycle 85. So, for a “difficult” case like the case of sample 3, a proper way to start the sR1 calculation is to find an optimal initial location for the starting single atom. This approach is general. It works even when no fragments of known structure are available for running the pR1 method to jumpstart the sR1 calculation. However, more study needs to be done on the effectiveness of this method.

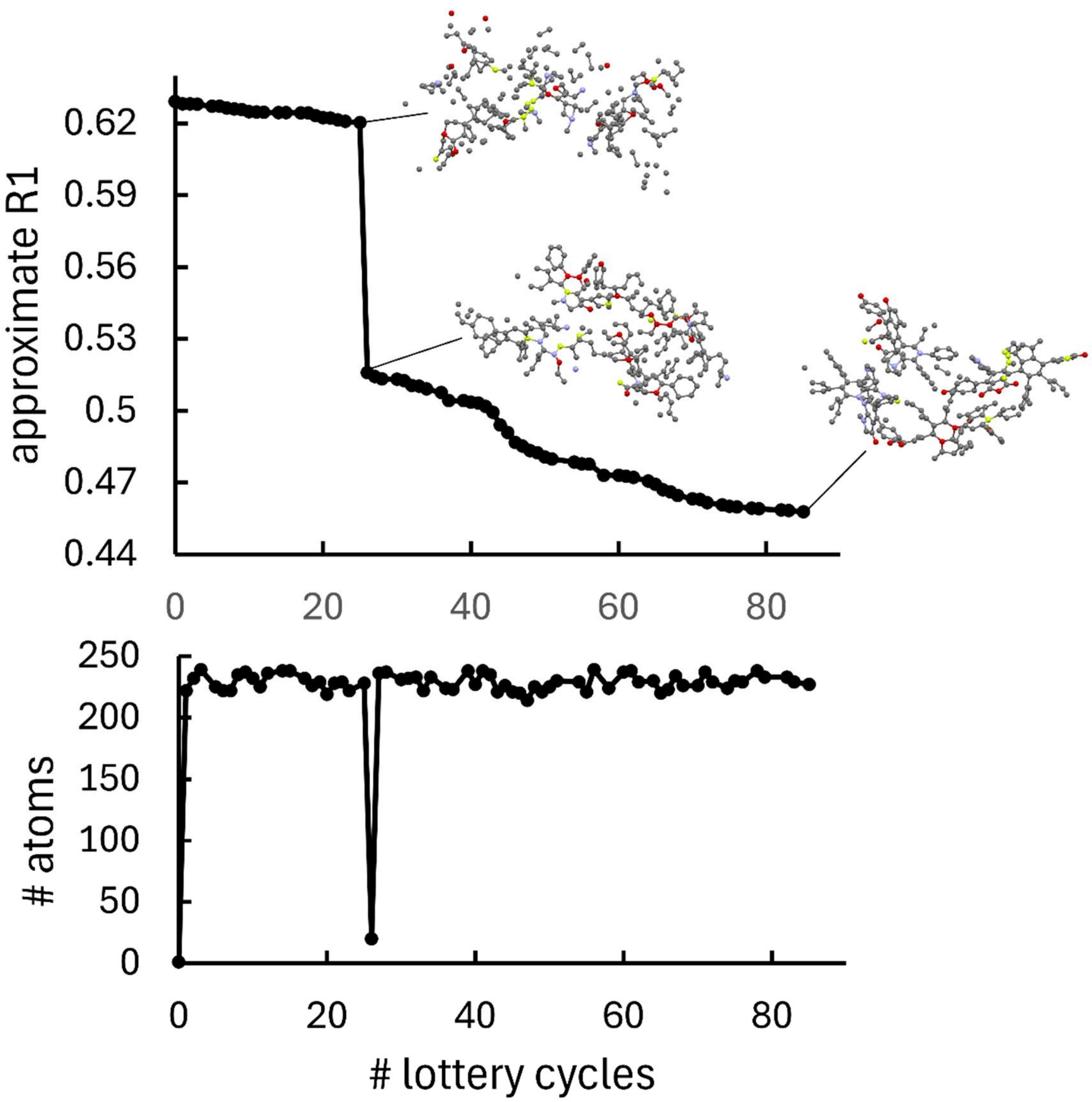

**Figure S5** Solving the structure of sample 3 by running the sR1 calculation in the lottery mode, starting from a single F atom at (0.325, 0.3042, 0.3). The top chart shows how the minimum approximate R1 that has been reached in each lottery cycle drops when the calculation proceeds. The bottom chart shows the size of the starting partial model of each lottery cycle. The inserts show the model structures at the end of cycles 25, 26, and 85, respectively.

Exact mathematical details are given in this paragraph: Let $a$, $b$, and $c$ be the size of the unit cell, and $s = 0.4$ Å be the step size of the grid. Calculate integers $N_x = int(a/s)$, $N_y = int(b/s)$, and $N_z = int(c/s)$. The grid points inside the cell are set at $x = i/N_x$, $y = j/N_y$, and $z = k/N_z$ for $i$ from 0 to $N_x - 1$, $j$ from 0 to $N_y - 1$, and $k$ from 0 to $N_z - 1$. Let $N_{cut} = 4$. Local grid points are set nearby point (0.3, 0.3, 0.3) at $x = 0.3 + i/N_x/N_{cut}$, $y = 0.3 + j/N_y/N_{cut}$, and $z = 0.3 + k/N_z/N_{cut}$ for $i$ from 0 to $N_{cut} - 1$, $j$ from 0 to $N_{cut} - 1$, and $k$ from 0 to $N_{cut} - 1$. There are total ${N_{cut}}^3 = 64$ local grid points. Perform one cycle of sR1 calculation by starting from a single F atom at each local grid point to find which local grid point can lead to an optimal tentative full model.

**S7. Solving the structure of sample 3 of a previous paper at low resolution**

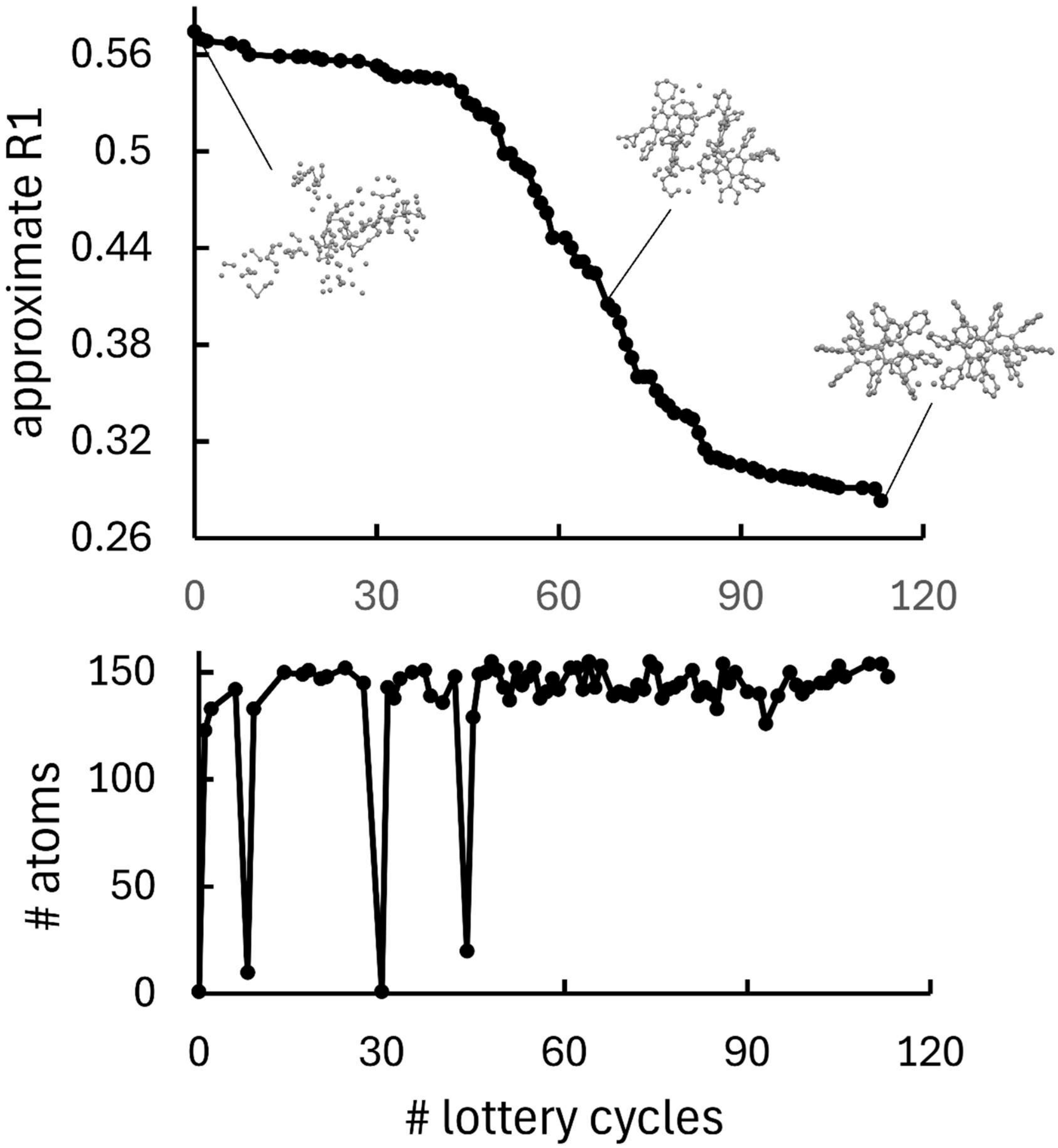

**Figure S6** Solving the structure of sample 3 of a previous paper (Zhang & Donahue, 2024) when data resolution was truncated to 1 Å by running the sR1 calculation in lottery mode, starting from a single C atom at (0.3, 0.3, 0.3). The top chart shows how the minimum approximate R1 in each lottery cycle drops when the calculation proceeds. The bottom chart shows the size of the starting partial model of each lottery cycle. The inserts show the model structures at the end of cycles 0, 68, and 113, respectively.

### S8. Using the Platon program to determine the space group and extracting the asymmetric unit

The sR1 method solves a structure in the minimal P1 space group. To find the correct space group, the result is checked by the *Platon* program (Spek, 2020). Using sample 3 as an example, the procedure is explained here. Checking the sR1 result of sample 3 the *Platon* gives following result:

```
Space Group  H-M:  P21/c                                      Laue:  2/m

Space Group Hall: -P 2ybc                          [Schoenflies: C2h^5 ]

Lattice Type: mP,  Centric,   Monoclinic, Multiplicity:   4( 2), No:  14


Non-Sohnke - No Absolute Structure


  Nr           ***** Symmetry Operation(s) *****


   1               X ,              Y ,              Z

   2             - X ,        1/2 + Y ,        1/2 - Z

   3             - X ,            - Y ,            - Z

   4               X ,        1/2 - Y ,        1/2 + Z


:: Origin Shifted to: 0.0985,-0.2459, 0.0497 after Transformation


::                                  (  0.0000  0.0000  1.0000) ( -0.0985)

:: R/t for Coordinates              (  0.0000 -1.0000  0.0000) (  0.2459)

::                                  (  1.0000  0.0000  0.0000) ( -0.0497)
```

So, the space group is $P2_1/c$. *Platon* also suggests how the coordinates should be transformed, including shift of origin to match the convention. Namely, new coordinates (x, y, z) are related to the old coordinates (x′, y′, z′) by x = z′ - 0.0985, y = -y′ + 0.2459, and z = x′ - 0.0497. Similarly, new hkl are related to the old h′k′l′ by h = l′, k = -k′, and l = h′. Convert xyz and hkl according to these rules. Also switch cell parameter a with c. Because there are four symmetry operators and four molecules in the unit cell, in this case, determination of the asymmetric unit is simple: just delete three molecules and only keep one molecule (any one is fine). Finally set up an input ".ins" file by including proper LATT and SYMM cards and assigning 0.05 for the initial isotropic displacement parameters. Using the SHELXL (Sheldrick, 2015b) program to perform LSQ refinement. After a few cycles of LSQ, R1 drops from initial 0.68 to 0.34 for 4419 reflections of Fo > 4sig(Fo). This intermediate model looks like:

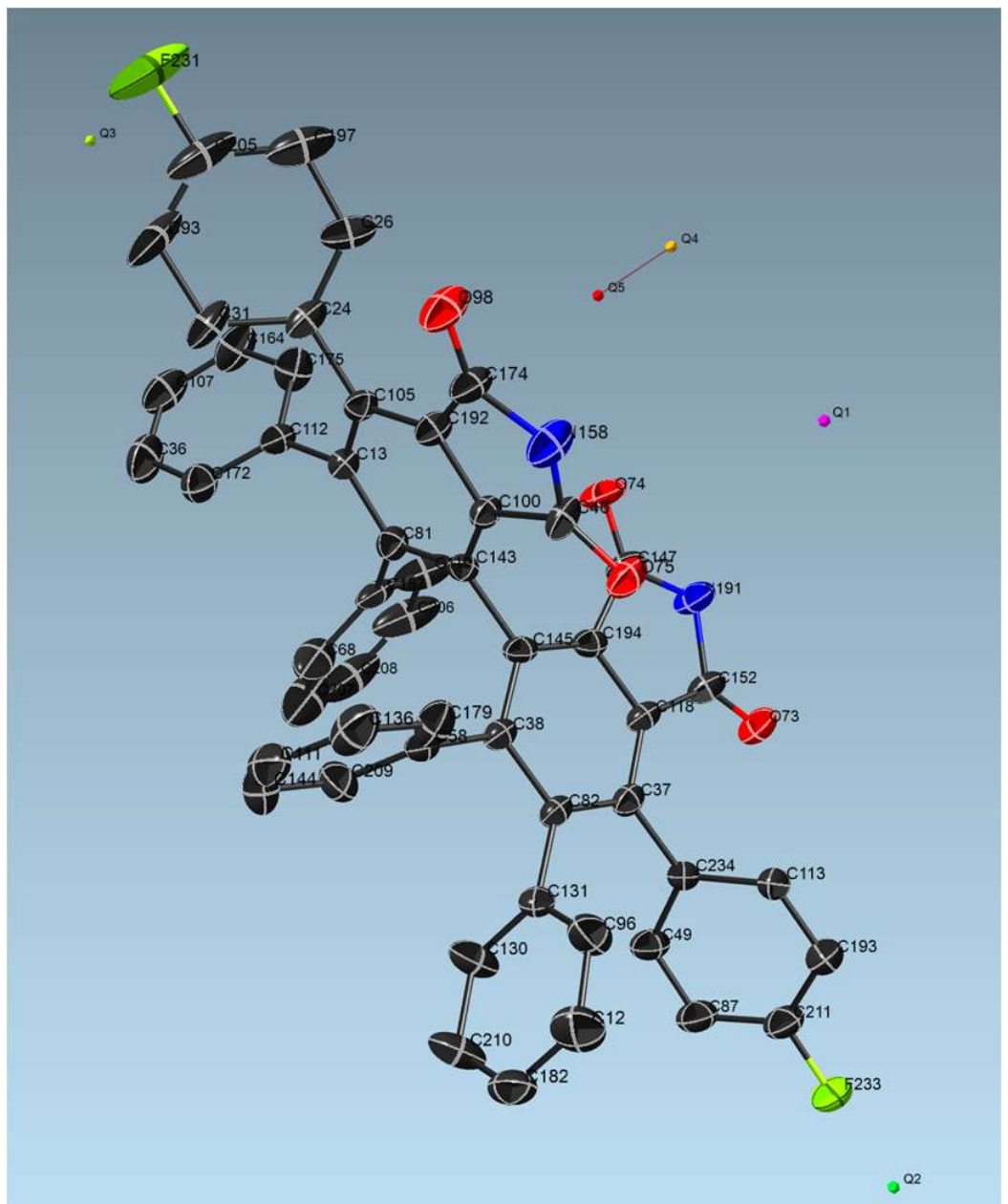

### S9. A list of samples that have been tested with the sR1 method

Note that the author adopts the convention of separating structure solution and structure refinement. In the author's daily work, two programs are used: the *SHELXT* program (Sheldrick, 2015a) and the *SHELXL* program (Sheldrick, 2015b). The result of *SHELXT* is called structure solution, which serves as input to *SHELXL*. The result of *SHELXL* is called structure refinement. When *SHELXT* yields a chemically sound and recognizable model, the structure solution is considered successful. To this point, at least the structure is qualitatively revealed. Depending on the data quality, a final refinement can or cannot be successful even if a structure is qualitatively solved. This same qualitative judgement is also applied to the sR1 method, namely, if the sR1 method produces chemically sound and recognizable model, it is regarded as having successfully solved the structure. Note that, for the four samples used in the body of this paper, as well as the three samples used in the previous paper (Zhang & Donahue, 2024), a quantitative comparison between the resulting model and the "correct" model was given. For that comparison, the "correct" model was determined in this way: the structure was solved by the *SHELXT* program and fully refined by the *SHELX*L program. After deleting the H atoms, the model is expanded into P1 space group. This expanded model is the "correct" model. Except for those seven samples, for all other samples such quantitative judgment was not used, instead, only the above-mentioned qualitative visual inspection was used. Of course, some datasets simply have no solution. However, only those datasets that can be solved by the *SHELXT* program were selected to test with sR1 method, so, all the selected datasets were solvable.

The following table lists datasets that have been tested by the sR1 method. Note that hydrogen atoms are omitted in the cell content and in the asymmetric unit. There are 224 datasets, among which 64 are light atoms only. Note that some datasets repeat the same compound when multiple datasets were acquired in a process of searching for a better quality. Some compounds also have crystals of different unit cells. The structures of these 224 datasets have been solved by the sR1 method, resulting in recognizable models. Most datasets were solved straightforwardly by starting from a single atom at (0.3, 0.3, 0.3). But there were 5 datasets (those labeled with letter “D”) appear to be difficult and required using the pR1 method to put some fragment(s) of a known structure to jumpstart the sR1 calculation. About 85 datasets (those labeled with letter “L”) have been tested with the lottery mode. For most datasets there was no need to test with the lottery mode, because the very first sR1 cycle already revealed a sizable recognizable partial structure, therefore, it was expected that the lottery mode would work trivially. The final R-values for reflections of Fo > 4 sig(Fo) have been listed in the same table. They were obtained after using *Platon* (Spek, 2020) to determine space group and asymmetric unit and performing refinement with *SHELXL* (Sheldrick, 2015a). Most datasets (184 datasets) have an R value smaller than 0.1. There are 22 cases having an R value between 0.1 and 0.16. These datasets are of intermediate quality. Finally, there are 18 cases having an R value higher than 0.16. These are low-quality datasets.

| Sample | R | Cell content | Asymmetric unit | a(Å) | b(Å) | c(Å) | α(°) | β(°) | $\gamma$(°) | space group |
|---|---|---|---|---|---|---|---|---|---|---|
| 1L | 0.022 | $(FN_2C_{13})_8$ | $FN_2C_{13}$ | 8.45 | 9.93 | 24.20 | 90.00 | 90.00 | 90.00 | Pbca |
| 2L | 0.24 | $(O_2C_{29})_8$ | $O_2C_{29}$ | 44.17 | 10.73 | 8.93 | 90.00 | 90.00 | 90.00 | Pbcn |
| 3L | 0.075 | $(F_2O_4N_2C_{52})_4$ | $F_2O_4N_2C_{52}$ | 12.27 | 24.23 | 15.42 | 90.00 | 96.94 | 90.00 | $P2_1/c$ |
| 4LD | 0.070 | $(PdS_6P_4C_{68})_4$ | $PdS_6P_4C_{68}$ | 23.32 | 14.45 | 22.39 | 90.00 | 115.22 | 90.00 | $P2_1/c$ |
| 5L | 0.021 | $(S_2O_2C_{12})_2$ | $S_2O_2C_{12}$ | 5.86 | 10.34 | 10.74 | 90.00 | 104.50 | 90.00 | $P2_1$ |
| 6L | 0.048 | $C_{46}$ | $C_{23}$ | 5.95 | 10.80 | 12.97 | 103.77 | 99.95 | 90.46 | P-1 |
| 7L | 0.040 | $(C_{78})_2$ | $C_{78}$ | 12.34 | 15.98 | 16.57 | 114.10 | 90.70 | 103.20 | P-1 |
| 8L | 0.020 | $(ION_2C_7)_4$ | $ION_2C_7$ | 18.34 | 7.18 | 8.26 | 90.00 | 90.00 | 90.00 | Pnma |
| 9 | 0.18 | $(PtP_2S_4N_4O_3C_{43})_2$ | $PtP_2S_4N_4O_3C_{43}$ | 11.60 | 14.16 | 14.99 | 68.04 | 68.02 | 74.14 | P-1 |
| 10 | 0.082 | $(MoSiO_4NC_{28})_8$ | $MoSiO_4NC_{28}$ | 36.36 | 10.22 | 20.00 | 90.00 | 111.34 | 90.00 | C2/c |
| 11 | 0.033 | $(IMo_3S_{13}N_3C_{27})_4$ | $IMo_3S_{13}N_3C_{27}$ | 12.97 | 37.99 | 11.64 | 90.00 | 91.72 | 90.00 | $P2_1/c$ |
| 12 | 0.059 | $(INC_8)_8$ | $INC_8$ | 23.13 | 8.34 | 14.88 | 90.00 | 126.57 | 90.00 | C2/c |
| 13 | 0.26 | $(MoC_{21})_6$ | $Mo_{1/3}C_7$ | 12.59 | 12.59 | 19.54 | 90.00 | 90.00 | 120.00 | R-3 |
| 14 | 0.034 | $(NiS_2N_2C_{18})_4$ | $NiS_2N_2C_{18}$ | 11.45 | 10.91 | 16.37 | 90.00 | 104.16 | 90.00 | $P2_1/n$ |
| 15 | 0.093 | $Pd_2Cl_4P_4O_2N_4C_{28}$ | $PdCl_2P_2ON_2C_{14}$ | 7.28 | 10.62 | 12.94 | 75.02 | 80.91 | 88.73 | P-1 |
| 16L | 0.068 | $(NiS_2N_2C_{18})_4$ | $NiS_2N_2C_{18}$ | 11.43 | 10.93 | 16.40 | 90.00 | 104.16 | 90.00 | $P2_1/n$ |
| 17L | 0.056 | $(S_8)_{16}$ | $S_4$ | 10.38 | 12.75 | 24.43 | 90.00 | 90.00 | 90.00 | Fddd |
| 18 | 0.27 | $(PtCl_4S_4C_{20})_{16}$ | $(PtCl_4S_4C_{20})_2$ | 24.48 | 18.82 | 29.59 | 90.00 | 101.18 | 90.00 | C2/c |
| 19 | 0.025 | $S_4P_2O_4C_{12}$ | $S_2PO_2C_6$ | 8.10 | 8.35 | 8.47 | 97.73 | 111.08 | 94.68 | P-1 |
| 20L | 0.071 | $(Cl_4O_2C_{14})_2$ | $Cl_2OC_7$ | 12.24 | 14.04 | 3.83 | 90.00 | 90.00 | 90.00 | Pba2 |
| 21L | 0.065 | $(NiCl_8S_4C_{28})_4$ | $NiCl_8S_4C_{28}$ | 8.25 | 15.58 | 23.86 | 90.00 | 90.00 | 94.69 | $P2_1/c$ |
| 22 | 0.040 | $(RePtBrS_2P_4O_4C_{82})_4$ | $(RePtBrS_2P_4O_4C_{82})_2$ | 12.59 | 23.36 | 25.24 | 105.83 | 103.93 | 96.80 | P-1 |
| 23 | 0.082 | $(IMo_6S_{26}P_6O_{12}C_{36})_4$ | $(IMo_6S_{26}P_6O_{12}C_{36})_2$ | 14.91 | 25.33 | 29.11 | 115.70 | 94.87 | 94.08 | P-1 |
| 24 | 0.040 | $(CuP_3F_6N_2C_{52})_4$ | $CuP_3F_6N_2C_{52}$ | 11.11 | 23.80 | 17.53 | 90.00 | 100.48 | 90.00 | $P2_1/n$ |
| 25 | 0.11 | $Mo_8Si_4O_{34}C_{60}$ | $Mo_4Si_2O_{17}C_{30}$ | 13.82 | 14.01 | 14.16 | 70.76 | 73.31 | 89.62 | P-1 |
| 26 | 0.036 | $(AgCl_2Si_2OC_{36})_8$ | $AgCl_2Si_2OC_{36}$ | 19.56 | 15.89 | 20.14 | 90.00 | 90.00 | 90.00 | Pbca |
| 27L | 0.057 | $S_6C_{24}$ | $S_3C_{12}$ | 9.14 | 9.61 | 9.90 | 72.06 | 71.95 | 63.88 | P-1 |
| 28 | 0.16 | $(PtS_6P_4C_{76})_8$ | $(PtS_6P_4C_{76})_4$ | 15.14 | 29.38 | 30.52 | 90.00 | 88.76 | 90.00 | P-1 |
| 29 | 0.034 | $(IMo_3S_{13}N_3C_{27})_4$ | $(IMo_3S_{13}N_3C_{27})_2$ | 11.87 | 37.77 | 13.14 | 90.00 | 92.34 | 90.00 | P-1 |
| 30 | 0.036 | $(AgCl_2Si_2OC_{36})_8$ | $AgCl_2Si_2OC_{36}$ | 19.55 | 15.89 | 20.15 | 90.00 | 90.00 | 90.00 | Pbca |
| 31 | 0.14 | $(IMo_3Se7S_6P_3O_6C_{18})_8$ | $(IMo_3Se7S_6P_3O_6C_{18})_2$ | 27.85 | 10.90 | 35.66 | 90.00 | 109.78 | 90.00 | $P2_1/n$ |
| 32 | 0.028 | $(IP_3OC_{52})_2$ | $IP_3OC_{52}$ | 9.80 | 18.49 | 12.97 | 90.00 | 103.33 | 90.00 | $P2_1$ |
| 33 | 0.029 | $(IP_3OC_{52})_2$ | $IP_3OC_{52}$ | 9.81 | 18.51 | 12.97 | 90.00 | 103.33 | 90.00 | $P2_1$ |
| 34 | 0.035 | $(Mo_2Se_6O_2N_2C_{18})_4$ | $Mo_2Se_6O_2N_2C_{18}$ | 11.33 | 17.93 | 14.56 | 90.00 | 95.77 | 90.00 | Cc |
| 35 | 0.12 | $Mo_6S_8P_6C_{12}$ | $Mo_3S_4P_3C_6$ | 11.84 | 11.98 | 12.07 | 101.76 | 115.33 | 97.00 | P-1 |
| 36 | 0.067 | $(Mo_3Se_7S_8P_4O_8C_{24})_2$ | $Mo_3Se_7S_8P_4O_8C_{24}$ | 13.76 | 14.36 | 16.19 | 72.26 | 72.91 | 68.91 | P-1 |
| 37 | 0.078 | $(Pd_2S_{12}C_{114})_4$ | $(Pd_2S_{12}C_{114})_2$ | 16.59 | 22.59 | 23.34 | 76.14 | 70.26 | 69.77 | P-1 |
| 38 | 0.042 | $(Mo_3Si_2O_{11}C_{36})_2$ | $Mo_3Si_2O_{11}C_{36}$ | 11.31 | 13.41 | 13.49 | 61.45 | 84.55 | 87.81 | P-1 |
| 39 | 0.068 | $(MoS_3Si_2O_2C_{36})_4$ | $(MoS_3Si_2O_2C_{36})_2$ | 11.12 | 13.24 | 23.89 | 90.00 | 96.73 | 90.00 | $P2_1$ |
| 40L | 0.11 | $(Cl_2O_2C_{14})_4$ | $Cl_2O_2C_{14}$ | 17.76 | 3.86 | 17.99 | 90.00 | 102.76 | 90.00 | P2/n |
| 41L | 0.046 | $(NiCl_8S_4C_{28})_4$ | $NiCl_8S_4C_{28}$ | 15.69 | 24.07 | 8.34 | 90.00 | 94.80 | 90.00 | $P2_1/c$ |
| 42L | 0.043 | $(NiCl_2S_2P_2C_{44})_2$ | $NiCl_2S_2P_2C_{44}$ | 12.11 | 12.11 | 14.63 | 67.01 | 79.65 | 73.17 | P-1 |
| 43 | 0.23 | $(Cl_4O_2C_{14})_4$ | $Cl_4O_2C_{14}$ | 18.59 | 3.82 | 18.59 | 90.00 | 97.96 | 90.00 | P2/m |
| 44L | 0.045 | $(Cl_4O_2C_{14})_2$ | $Cl_2OC_7$ | 12.18 | 14.04 | 3.83 | 90.00 | 90.00 | 90.00 | Pba2 |

| Sample | R | Cell content | Asymmetric unit | a(Å) | b(Å) | c(Å) | α(°) | β(°) | $\gamma$(°) | space group |
|---|---|---|---|---|---|---|---|---|---|---|
| 45 | 0.045 | $(SO_3NC_{13})_2$ | $(SO_3NC_{13})_2$ | 7.44 | 7.42 | 10.60 | 89.98 | 75.82 | 89.98 | P1 |
| 46 | 0.046 | $(SN_2C_{16})_2$ | $SN_2C_{16}$ | 6.96 | 9.00 | 12.25 | 92.74 | 95.60 | 106.98 | P-1 |
| 47 | 0.051 | $(SO_2N_2C_{26})_4$ | $(SO_2N_2C_{26})_2$ | 12.17 | 13.18 | 14.58 | 91.76 | 107.65 | 98.31 | P-1 |
| 48 | 0.18 | $(CuS_2P_2F_{12}N_8C_{18})_2$ | $Cu_{1/2}SPF_6N_4C_9$ | 10.93 | 14.67 | 8.14 | 90.00 | 90.00 | 90.00 | $P2_12_12$ |
| 49 | 0.050 | $(C_{32})_4$ | $C_{32}$ | 11.27 | 14.80 | 15.78 | 90.00 | 97.94 | 90.00 | $P2_1/n$ |
| 50 | 0.057 | $(W_2S_8N_2C_{18})_4$ | $W_2S_8N_2C_{18}$ | 19.90 | 9.74 | 16.68 | 90.00 | 112.11 | 90.00 | $P2_1/c$ |
| 51 | 0.041 | $(C_{19}ON_2)_4$ | $C_{19}ON_2$ | 10.84 | 17.82 | 9.55 | 90.00 | 94.39 | 90.00 | $P2_1/c$ |
| 52 | 0.041 | $(C_{36}S_2)_2$ | $C_{18}S$ | 6.30 | 15.58 | 14.80 | 90.00 | 102.11 | 90.00 | $P2_1/c$ |
| 53 | 0.042 | $(C_{30}F_2)_2$ | $C_{15}F$ | 5.82 | 11.97 | 15.17 | 90.00 | 94.03 | 90.00 | $P2_1/c$ |
| 54 | 0.050 | $(W_3IS_{13}N_3C_{37})_4$ | $W_3IS_{13}N_3C_{37}$ | 12.97 | 38.01 | 11.75 | 90.00 | 91.46 | 90.00 | $P2_1/c$ |
| 55 | 0.096 | $(PtCl_4P_2C_{28})_8$ | $(PtCl_4P_2C_{28})_2$ | 18.09 | 18.34 | 16.41 | 90.00 | 90.00 | 90.00 | $P2_1/c$ |
| 56 | 0.23 | $(F_4C_{18})_4$ | $F_2C_9$ | 5.77 | 28.74 | 7.62 | 90.00 | 90.00 | 90.00 | $C222_1$ |
| 57 | 0.33 | $(Ni_6S_{12}C_{84})_8$ | $Ni_6S_{12}C_{84}$ | 26.05 | 26.23 | 26.38 | 90.00 | 90.00 | 90.00 | Pbca |
| 58 | 0.098 | $(Ni_6S_{12}C_{84})_8$ | $Ni_6S_{12}C_{84}$ | 25.99 | 26.18 | 26.35 | 90.00 | 90.00 | 90.00 | Pbca |
| 59 | 0.30 | $(KCl_2O_{12}N_4C_{12})_3$ | $(KCl_2O_{12}N_4C_{12})_3$ | 10.81 | 10.83 | 11.92 | 82.07 | 85.23 | 67.92 | P1 |
| 60 | 0.058 | $(KS_2P2O_7NC_{14})_4$ | $KS_2P2O_7NC_{14}$ | 11.72 | 17.50 | 13.06 | 90.00 | 114.12 | 90.00 | $P2_1/c$ |
| 61 | 0.12 | $(KClO_5N_2C_6)_6$ | $(KClO_5N_2C_6)_3$ | 10.87 | 10.89 | 11.94 | 82.20 | 85.03 | 67.60 | P-1 |
| 62 | 0.36 | $(Pd_2Cl_5S_6P_4C_{105})_4$ | $(PtPdCl_5S_6P_4C_{105})_2$ | 16.47 | 23.10 | 31.39 | 71.40 | 90.00 | 90.00 | P-1 |
| 63 | 0.037 | $(W_3BrS_{13}N_3C_{33})_4$ | $W_3BrS_{13}N_3C_{33}$ | 13.23 | 37.62 | 11.71 | 90.00 | 93.21 | 90.00 | $P2_1/c$ |
| 64L | 0.068 | $(NiCl_4S_2P_2C_{45})_4$ | $NiCl_4S_2P_2C_{45}$ | 13.90 | 19.39 | 15.06 | 90.00 | 93.68 | 90.00 | $P2_1/c$ |
| 65 | 0.29 | $(NiS_2P_2O_6C_{45})_4$ | $(NiS_2P_2O_6C_{45})_2$ | 17.85 | 18.15 | 18.79 | 110.13 | 109.75 | 92.46 | P-1 |
| 66L | 0.21 | $(S_4C_{30})_4$ | $S_4C_{30}$ | 11.86 | 20.68 | 11.77 | 90.00 | 90.50 | 90.00 | $P2_1/c$ |
| 67 | 0.032 | $(P_2O_6NC_4)_4$ | $P_2O_6NC_4$ | 7.95 | 10.19 | 11.75 | 90.00 | 91.79 | 90.00 | Cc |
| 68 | 0.28 | $(PtPdCl_8S_6P_4C_{104})_4$ | $PtPdCl_8S_6P_4C_{104}$ | 22.85 | 17.13 | 31.68 | 90.00 | 107.09 | 90.00 | $P2_1/c$ |
| 69 | 0.045 | $(Ni_6CoS_{12}NC_{106})_8$ | $Ni_6CoS_{12}NC_{106}$ | 25.95 | 18.38 | 43.06 | 90.00 | 94.06 | 90.00 | C2/c |
| 70 | 0.066 | $(Mo_3S_{12}ON_5C_{27})_4$ | $Mo_3S_{12}ON_5C_{27}$ | 21.87 | 10.34 | 19.77 | 90.00 | 94.80 | 90.00 | $P2_1/c$ |
| 71 | 0.31 | $(SP_4O_{12}C_{27})_2$ | $SP_4O_{12}C_{27}$ | 12.21 | 12.22 | 17.30 | 104.50 | 96.39 | 118.79 | P-1 |
| 72 | 0.048 | $(IMo_3Se_7S_6N_3C_{27})_{12}$ | $(IMo_3Se_7S_6N_3C_{27})_3$ | 27.76 | 13.57 | 40.82 | 90.00 | 91.02 | 90.00 | $P2_1/c$ |
| 73 | 0.25 | $(IMo_3Se_3S_{10}N_3C_{27})_{12}$ | $(IMo_3Se_3S_{10}N_3C_{27})_3$ | 27.75 | 13.64 | 46.36 | 90.00 | 101.69 | 90.00 | $P2_1/n$ |
| 74 | 0.061 | $(IN_3C_5)_4$ | $(IN_3C_5)_2$ | 10.05 | 10.04 | 10.96 | 117.07 | 117.04 | 90.42 | P-1 |
| 75 | 0.047 | $(ClSiO_3C_{17})_8$ | $(ClSiO_3C_{17})_2$ | 17.41 | 9.48 | 22.74 | 90.00 | 92.53 | 90.00 | $P2_1/n$ |
| 76 | 0.053 | $(O_2N_5C_{25})_4$ | $O_2N_5C_{25}$ | 18.61 | 4.98 | 22.69 | 90.00 | 99.48 | 90.00 | $P2_1/n$ |
| 77 | 0.032 | $(ClSO_4NC_{16})_8$ | $ClSO_4NC_{16}$ | 9.57 | 15.01 | 21.98 | 90.00 | 90.00 | 90.00 | Pbca |
| 78 | 0.049 | $(O_2N_3C_{25})_4$ | $O_2N_3C_{25}$ | 17.49 | 7.09 | 17.65 | 90.00 | 105.77 | 90.00 | $P2_1/c$ |
| 79 | 0.043 | $(BrO_2N_4C_{33})_4$ | $BrO_2N_4C_{33}$ | 13.12 | 14.54 | 15.56 | 90.00 | 107.12 | 9.00 | $P2_1/n$ |
| 80 | 0.087 | $(O_6C_{13})_8$ | $(O_6C_{13})_2$ | 7.12 | 9.64 | 39.66 | 90.00 | 90.00 | 90.00 | $P2_12_12_1$ |
| 81L | 0.038 | $(FN_2C_{13})_8$ | $FN_2C_{13}$ | 8.45 | 9.94 | 24.20 | 90.00 | 90.00 | 90.00 | Pbca |
| 82LD | 0.10 | $(OC_{30})_8$ | $OC_{30}$ | 44.05 | 10.69 | 8.89 | 90.00 | 90.00 | 90.00 | Pbcn |
| 83D | 0.15 | $(OC_{46})_8$ | $(OC_{46})_4$ | 13.87 | 16.59 | 36.17 | 90.00 | 90.00 | 94.51 | P-1 |
| 84 | 0.082 | $(OC_{52})_8$ | $(OC_{52})_2$ | 13.91 | 35.89 | 16.66 | 90.00 | 94.47 | 90.00 | $P2_1/n$ |
| 85D | 0.15 | $(C_{52})_8$ | $(C_{52})_2$ | 13.71 | 36.11 | 17.20 | 90.00 | 94.54 | 90.00 | $P2_1/n$ |
| 86 | 0.067 | $(O_3C_{29})_8$ | $O_3C_{29}$ | 22.04 | 24.62 | 7.85 | 90.00 | 90.00 | 90.00 | Pccn |
| 87 | 0.12 | $(BrOC_{52})_4$ | $(BrOC_{52})_2$ | 10.90 | 19.50 | 21.65 | 80.36 | 75.79 | 81.10 | P-1 |
| 88 | 0.11 | $(BrOC_{52})_4$ | $(BrOC_{52})_2$ | 10.90 | 10.92 | 17.51 | 73.24 | 85.95 | 76.13 | P-1 |

| Sample | R | Cell content | Asymmetric unit | a(Å) | b(Å) | c(Å) | α(°) | β(°) | $\gamma$(°) | space group |
|---|---|---|---|---|---|---|---|---|---|---|
| 89D | 0.21 | $(C_{52})_6$ | $(C_{52})_3$ | 12.56 | 12.56 | 36.32 | 90.40 | 90.40 | 98.46 | P-1 |
| 90 | 0.042 | $(ClO_{11}N_7C_{26})_2$ | $ClO_{11}N_7C_{26}$ | 8.95 | 12.19 | 14.88 | 83.86 | 75.60 | 80.45 | P-1 |
| 91 | 0.27 | $(ClO_5N_7C_{26})_2$ | $ClO_5N_7C_{26}$ | 8.43 | 13.42 | 13.68 | 108.22 | 103.75 | 107.49 | P-1 |
| 92 | 0.041 | $(PtO_3N_5C_{28})_8$ | $PtO_3N_5C_{28}$ | 23.32 | 14.45 | 22.39 | 90.00 | 115.22 | 90.00 | Aba2 |
| 93 | 0.046 | $(ON_3C_{20})_8$ | $ON_3C_{20}$ | 29.91 | 4.99 | 25.88 | 90.00 | 125.14 | 90.00 | C2/c |
| 94 | 0.066 | $(ONC_{15})_{16}$ | $(ONC_{15})_2$ | 19.87 | 25.52 | 9.52 | 90.00 | 90.00 | 90.00 | Pccn |
| 95 | 0.12 | $(ON_3C_{16})_4$ | $ON_3C_{16}$ | 15.93 | 4.79 | 20.43 | 90.00 | 111.82 | 90.00 | P2/c |
| 96 | 0.045 | $(O_2C_4)_4$ | $OC_2$ | 15.12 | 3.90 | 15.84 | 90.00 | 106.49 | 90.00 | C2/c |
| 97 | 0.052 | $(ON_3C_{21})_8$ | $ON_3C_{21}$ | 22.47 | 22.47 | 6.48 | 90.00 | 90.00 | 90.00 | $P4_2/n$ |
| 98 | 0.088 | $(O_3N_3C_{16})_4$ | $O_3N_3C_{16}$ | 15.89 | 4.78 | 20.36 | 90.00 | 111.87 | 90.00 | P2/c |
| 99 | 0.046 | $(SN_2C_{18})_2$ | $SN_2C_{18}$ | 6.96 | 9.00 | 12.25 | 92.74 | 95.60 | 106.98 | P-1 |
| 100 | 0.051 | $(SO_2N_2C_{26})_4$ | $(SO_2N_2C_{26})_2$ | 12.17 | 13.18 | 14.58 | 91.76 | 107.65 | 98.31 | P-1 |
| 101 | 0.039 | $(O_3N_4C_{28})_4$ | $O_3N_4C_{28}$ | 17.33 | 17.34 | 7.78 | 90.00 | 90.00 | 90.00 | $Pna2_1$ |
| 102 | 0.10 | $(O_3N_2C_{21})_4$ | $(O_3N_2C_{21})_4$ | 9.57 | 11.84 | 16.03 | 89.98 | 90.10 | 90.04 | P1 |
| 103 | 0.10 | $(ON_3C_{22})_4$ | $(ON_3C_{22})_2$ | 5.86 | 9.62 | 30.72 | 90.00 | 90.00 | 93.68 | P-1 |
| 104L | 0.070 | $(PdS_6P_4C_{68})_4$ | $PdS_6P_4C_{68}$ | 23.32 | 14.45 | 22.39 | 90.00 | 115.22 | 90.00 | $P2_1/c$ |
| 105L | 0.055 | $(S_4O_2C_{20})_2$ | $S_4O_2C_{20}$ | 10.38 | 12.75 | 24.43 | 90.00 | 90.00 | 90.00 | Pc |
| 106L | 0.054 | $Fe_3Cl_6N_2C_{82}$ | $Fe_{3/2}Cl3NC_{41}$ | 10.25 | 13.25 | 15.16 | 70.13 | 76.32 | 87.71 | P-1 |
| 107L | 0.058 | $(P_2C_{26})_4$ | $P_2C_{26}$ | 12.76 | 12.99 | 14.07 | 90.00 | 116.69 | 90.00 | $P2_1/c$ |
| 108L | 0.095 | $(ReI_3S_4N_2C_{18})_2$ | $ReI_3S_4N_2C_{18}$ | 8.21 | 9.73 | 19.92 | 76.01 | 78.11 | 65.04 | P-1 |
| 109L | 0.047 | $(OC_{15})_4$ | $OC_{15}$ | 4.14 | 13.99 | 17.12 | 90.00 | 90.00 | 90.00 | $P2_12_12_1$ |
| 110L | 0.036 | $(O_2C_{30})_2$ | $O_2C_{30}$ | 9.54 | 10.79 | 11.20 | 74.42 | 65.97 | 76.35 | P-1 |
| 111L | 0.024 | $(S_4P_2C_{16})_4$ | $S_4P_2C_{16}$ | 11.55 | 18.17 | 11.79 | 90.00 | 107.95 | 90.00 | $P2_1/c$ |
| 112L | 0.020 | $(BrC_{14})_4$ | $BrC_{14}$ | 3.97 | 13.99 | 18.33 | 90.00 | 95.94 | 90.00 | $P2_1/c$ |
| 113L | 0.055 | $(O_2C_{30})_4$ | $O_2C_{30}$ | 12.64 | 6.25 | 25.05 | 90.00 | 90.00 | 90.00 | $Pna2_1$ |
| 114L | 0.10 | $(S_2NC_{13})_4$ | $S_2NC_{13}$ | 7.42 | 14.21 | 12.59 | 90.00 | 106.50 | 90.00 | $P2_1/n$ |
| 115L | 0.023 | $(BrC_{14})_4$ | $BrC_{14}$ | 3.97 | 14.00 | 18.33 | 90.00 | 95.95 | 90.00 | $P2_1/c$ |
| 116L | 0.048 | $(O_2NC_7)_2$ | $O_2NC_7$ | 7.33 | 7.45 | 7.81 | 96.02 | 108.53 | 117.62 | P-1 |
| 117L | 0.061 | $(BrO_3N_3C_{15})_8$ | $BrO_3N_3C_{15}$ | 25.97 | 10.55 | 12.38 | 90.00 | 101.24 | 90.00 | C2/c |
| 118L | 0.025 | $(BrO_3C_{11})_4$ | $BrO_3C_{11}$ | 7.09 | 14.39 | 10.04 | 90.00 | 90.00 | 90.00 | $Pna2_1$ |
| 119L | 0.054 | $(SF_3O_4N_2C_{14})_4$ | $(SF_3O_4N_2C_{14})_2$ | 4.63 | 14.46 | 23.63 | 94.02 | 93.11 | 96.48 | P-1 |
| 120L | 0.038 | $(O_5C_{13})_8$ | $O_5C_{13}$ | 8.22 | 13.87 | 20.84 | 90.00 | 96.06 | 90.00 | C2/c |
| 121L | 0.12 | $(I_2O_2NC_{14})_8$ | $I_2O_2NC_{14}$ | 37.75 | 4.52 | 17.33 | 90.00 | 116.74 | 90.00 | C2/c |
| 122L | 0.026 | $(I_2O_4NC_{16})_4$ | $I_2O_4NC_{16}$ | 7.91 | 14.70 | 14.07 | 90.00 | 90.00 | 90.00 | $Pca2_1$ |
| 123L | 0.029 | $(BrOC_{50})_2$ | $BrOC_{50}$ | 12.97 | 13.06 | 13.41 | 70.14 | 82.72 | 60.48 | P-1 |
| 124L | 0.071 | $(BrC_{13})_4$ | $BrC_{13}$ | 10.21 | 10.96 | 11.54 | 90 | 112.21 | 90 | $P2_1/c$ |
| 125L | 0.045 | $(ON_2C_{21})_4$ | $ON_2C_{21}$ | 10.31 | 11.60 | 12.15 | 90.00 | 99.04 | 90.00 | Cc |
| 126 | 0.031 | $(SO_3NC_{21})_4$ | $SO_3NC_{21}$ | 6.16 | 12.07 | 23.42 | 90.00 | 96.31 | 90.00 | $P2_1/c$ |
| 127 | 0.029 | $(SFO_3NC_{15})_4$ | $SFO_3NC_{15}$ | 11.83 | 8.29 | 14.16 | 90.00 | 102.81 | 90.00 | $P2_1/c$ |
| 128 | 0.072 | $(SO_5NC_{12})_4$ | $SO_5NC_{12}$ | 8.27 | 8.81 | 17.32 | 90.00 | 98.85 | 90.00 | $P2_1/n$ |
| 129 | 0.046 | $(SO_5NC_{12})_4$ | $SO_5NC_{12}$ | 8.27 | 8.81 | 17.33 | 90.00 | 98.83 | 90.00 | $P2_1/n$ |
| 130 | 0.028 | $(SCl_2O_3NC_{15})_4$ | $SCl_2O_3NC_{15}$ | 13.57 | 11.74 | 9.48 | 90.00 | 101.05 | 90.00 | $P2_1/c$ |
| 131 | 0.030 | $(SClO_3NC_{15})_4$ | $SClO_3NC_{15}$ | 13.38 | 11.17 | 9.77 | 90.00 | 98.65 | 90.00 | $P2_1/c$ |
| 132 | 0.032 | $(SClO_3NC_{15})_4$ | $SClO_3NC_{15}$ | 13.38 | 11.16 | 9.78 | 90.00 | 98.62 | 90.00 | $P2_1/c$ |

| Sample | R | Cell content | Asymmetric unit | a(Å) | b(Å) | c(Å) | α(°) | β(°) | $\gamma$(°) | space group |
|---|---|---|---|---|---|---|---|---|---|---|
| 133 | 0.039 | $(SO_3NC_{16})_4$ | $(SO_3NC_{16})_2$ | 9.85 | 11.81 | 15.65 | 104.33 | 101.29 | 97.68 | P-1 |
| 134 | 0.039 | $(SO_3NC_{21})_4$ | $SO_3NC_{21}$ | 5.98 | 12.06 | 25.18 | 90.00 | 90.00 | 90.00 | $P2_12_12_1$ |
| 135 | 0.034 | $(SO_5NC_{11})_4$ | $SO_5NC_{11}$ | 12.67 | 7.75 | 12.97 | 90.00 | 95.86 | 90.00 | $P2_1/n$ |
| 136 | 0.078 | $(SO_5NC_{18})_4$ | $SO_5NC_{18}$ | 10.43 | 10.49 | 15.83 | 90.00 | 99.98 | 90.00 | $P2_1/c$ |
| 137L | 0.15 | $(SO_5NC_{17})_8$ | $(SO_5NC_{17})_2$ | 21.34 | 10.24 | 16.49 | 90.00 | 96.82 | 90.00 | $P2_1/c$ |
| 138 | 0.063 | $(SO_5NC_{17})_8$ | $(SO_5NC_{17})_2$ | 21.32 | 10.24 | 16.49 | 90.00 | 96.80 | 90.00 | $P2_1/c$ |
| 139 | 0.044 | $(SO_3NC_{17})_8$ | $(SO_3NC_{17})_2$ | 21.33 | 11.62 | 11.90 | 90.00 | 90.00 | 90.00 | $Pna2_1$ |
| 140 | 0.031 | $(SO_3N_3C_{11})_8$ | $(SO_3N_3C_{11})_2$ | 26.39 | 10.45 | 9.02 | 90.00 | 96.33 | 90.00 | $P2_1/c$ |
| 141 | 0.025 | $(NiS_4C_{36})_2$ | $Ni_{1/2}S_2C_{18}$ | 8.44 | 14.09 | 13.59 | 90.00 | 96.15 | 90.00 | $P2_1/c$ |
| 142 | 0.058 | $(P_2C_{26})_4$ | $P_2C_{26}$ | 12.76 | 12.99 | 14.07 | 90.00 | 116.69 | 90.00 | $P2_1/c$ |
| 143 | 0.072 | $(Sn_2S_8N_2C_{18})_4$ | $Sn_2S_8N_2C_{18}$ | 18.98 | 9.70 | 16.74 | 90.00 | 111.68 | 90.00 | $P2_1/c$ |
| 144 | 0.018 | $(SnS_2N_2C_6)_4$ | $SnS_2N_2C_6$ | 12.34 | 7.80 | 9.75 | 90.00 | 90.00 | 90.00 | Pnnm |
| 145 | 0.059 | $(S_{13}F_{36}N_6C_{30})_2$ | $S_{13}F_{36}N_6C_{30}$ | 12.09 | 12.40 | 20.28 | 75.84 | 75.22 | 88.40 | P-1 |
| 146L | 0.053 | $(Se_5N_2C_{10})_4$ | $Se_5N_2C_{10}$ | 6.68 | 9.91 | 25.36 | 90.00 | 90.00 | 90.00 | P-1 |
| 147 | 0.052 | $(Mo_3ISe_{13}N3C_{27})_{12}$ | $(Mo_3ISe_{13}N3C_{27})_3$ | 28.18 | 13.74 | 41.02 | 90.00 | 90.92 | 90.00 | $P2_1/c$ |
| 148L | 0.064 | $(O_2C_{16})_4$ | $(O_2C_{16})_2$ | 7.55 | 14.00 | 11.97 | 90.00 | 94.93 | 90.00 | $P2_1$ |
| 149 | 0.024 | $(SO_2C_{11})_4$ | $SO_2C_{11}$ | 5.88 | 8.96 | 18.36 | 90.00 | 90.00 | 90.00 | $P2_12_12_1$ |
| 150 | 0.066 | $(O_3C_{31})_4$ | $O_3C_{31}$ | 9.14 | 10.88 | 15.43 | 88.55 | 74.15 | 84.69 | P-1 |
| 151 | 0.15 | $(Mo_3S_{12}N_5C_{39})_4$ | $(Mo_3S_{12}N_5C_{39})_2$ | 11.23 | 18.28 | 32.61 | 90.00 | 99.76 | 90.00 | Pn |
| 152 | 0.035 | $(Mo_3S_{12}ON_5C_{23})_2$ | $Mo_3S_{12}ON_5C_{23}$ | 10.61 | 11.40 | 19.04 | 87.49 | 80.55 | 69.72 | P-1 |
| 153 | 0.34 | $(Mo_3BrS_{13}P_6N_3O_{18}C_{36})_8$ | $(Mo_3BrS_{13}P_6N_3O_{18}C_{36})_2$ | 33.69 | 12.51 | 36.59 | 90.00 | 98.76 | 90.00 | $P2_1/n$ |
| 154 | 0.098 | $(NiS_4C_{60})_4$ | $Ni_{1/2}S_2C_{30}$ | 25.70 | 14.67 | 19.27 | 90.00 | 116.92 | 90.00 | C2/c |
| 155 | 0.079 | $(Re_2Se_4O_6N_2C_{11})_2$ | $Re_2Se_4O_6N_2C_{11}$ | 11.46 | 10.70 | 12.45 | 84.76 | 81.72 | 63.17 | P-1 |
| 156 | 0.036 | $(ON_3C_{11})_4$ | $ON_3C_{11}$ | 10.93 | 18.43 | 4.75 | 90.00 | 90.00 | 90.00 | $Pna2_1$ |
| 157 | 0.029 | $(ClON_4C_{16})_4$ | $ClON_4C_{16}$ | 13.30 | 13.21 | 8.63 | 90.00 | 106.20 | 90.00 | $P2_1/c$ |
| 158L | 0.073 | $(O_2N_4C_{38})_4$ | $O_2N_4C_{38}$ | 17.36 | 10.89 | 17.92 | 90.00 | 117.32 | 90.00 | $P2_1/c$ |
| 159 | 0.040 | $(FO_4N_3C_{18})_2$ | $FO_4N_3C_{18}$ | 8.33 | 9.47 | 10.45 | 102.48 | 96.75 | 96.84 | P-1 |
| 160L | 0.038 | $(O_2N_4C_{11})_4$ | $O_2N_4C_{11}$ | 10.89 | 9.25 | 11.71 | 90.00 | 115.35 | 90.00 | $P2_1/c$ |
| 161 | 0.034 | $(SO_6N_2C_{16})_2$ | $SO_6N_2C_{16}$ | 8.85 | 10.69 | 12.71 | 74.94 | 83.33 | 84.22 | P-1 |
| 162 | 0.032 | $(ClSO_4NC_{16})_8$ | $ClSO_4NC_{16}$ | 9.57 | 15.01 | 21.98 | 90.00 | 90.00 | 90.00 | Pbca |
| 163 | 0.23 | $(SO_4NC_{17})_4$ | $(SO_4NC_{17})_2$ | 11.77 | 5.57 | 25.09 | 90.00 | 103.17 | 90.00 | $P2_1$ |
| 164 | 0.14 | $(SO_4NC_{17})_4$ | $SO_4NC_{17}$ | 11.76 | 5.57 | 25.03 | 90.00 | 102.77 | 90.00 | $P2_1/n$ |
| 165 | 0.042 | $(SO_4NC_{17})_4$ | $SO_4NC_{17}$ | 11.76 | 5.57 | 25.03 | 90.00 | 102.73 | 90.00 | $P2_1/n$ |
| 166L | 0.046 | $(O_5C_{10})_2$ | $O_5C_{10}$ | 7.80 | 8.53 | 8.57 | 100.06 | 95.46 | 95.26 | P-1 |
| 167L | 0.035 | $(O_3N_2C_7)_4$ | $O_3N_2C_7$ | 16.04 | 6.40 | 7.16 | 90.00 | 90.00 | 90.00 | Pnma |
| 168 | 0.047 | $(ClSO_4NC_{21})_4$ | $ClSO_4NC_{21}$ | 8.07 | 19.00 | 13.93 | 90.00 | 92.33 | 90.00 | $P2_1/n$ |
| 169 | 0.030 | $(SO_3N_5C_8)_4$ | $SO_3N_5C_8$ | 4.85 | 14.39 | 15.56 | 90.00 | 91.80 | 90.00 | $P2_1/c$ |
| 170 | 0.037 | $(SO_5N_3C_{24})_4$ | $SO_5N_3C_{24}$ | 9.17 | 19.63 | 13.02 | 90.00 | 91.01 | 90.00 | $P2_1/n$ |
| 171 | 0.049 | $(O_2N_3C_{25})_4$ | $O_2N_3C_{25}$ | 17.49 | 7.09 | 17.65 | 90.00 | 105.77 | 90.00 | $P2_1/c$ |
| 172 | 0.064 | $(ClO_2N_4C_{31})_4$ | $ClO_2N_4C_{31}$ | 12.86 | 4.54 | 15.68 | 90.00 | 106.67 | 90.00 | $P2_1/n$ |
| 173 | 0.043 | $(BrO_2N_4C_{33})_4$ | $BrO_2N_4C_{33}$ | 13.11 | 14.54 | 15.60 | 90.00 | 107.12 | 90.00 | $P2_1/n$ |
| 174L | 0.030 | $(SO_3NC_6)_4$ | $SO_3NC_6$ | 9.49 | 6.92 | 11.72 | 90.00 | 103.29 | 90.00 | $P2_1/c$ |
| 175 | 0.039 | $(SO_5N_3C_{20})_2$ | $SO_5N_3C_{20}$ | 8.19 | 9.96 | 12.59 | 89.22 | 81.73 | 75.95 | P-1 |
| 176 | 0.035 | $(Cl_2SO_4N_3C_{23})_4$ | $Cl_2SO_4N_3C_{23}$ | 9.17 | 20.25 | 13.12 | 90.00 | 91.34 | 90.00 | $P2_1/n$ |

| Sample | R | Cell content | Asymmetric unit | a(Å) | b(Å) | c(Å) | α(°) | β(°) | $\gamma$(°) | space group |
|---|---|---|---|---|---|---|---|---|---|---|
| 177L | 0.17 | $(SO_5N_4C_{18})_8$ | $(SO_5N_4C_{18})_4$ | 8.13 | 16.27 | 28.38 | 93.20 | 93.14 | 102.09 | P-1 |
| 178 | 0.039 | $(SO_2N_2C_{18})_2$ | $SO_2N_2C_{18}$ | 6.87 | 9.26 | 13.88 | 106.35 | 93.10 | 107.42 | P-1 |
| 179 | 0.059 | $(Cl_2SO_4N_2C_{16})_4$ | $Cl_2SO_4N_2C_{16}$ | 8.14 | 9.35 | 22.27 | 90.00 | 96.38 | 90.00 | $P2_1/n$ |
| 180 | 0.030 | $(S_4O_8N_6C_{38})_2$ | $S_4O_8N_6C_{38}$ | 8.24 | 15.04 | 15.87 | 101.68 | 104.67 | 98.86 | P-1 |
| 181 | 0.052 | $(S_2O_5N_3C_{19})_4$ | $S_2O_5N_3C_{19}$ | 7.53 | 12.47 | 20.54 | 90.00 | 92.47 | 90.00 | $P2_1/c$ |
| 182 | 0.042 | $(ClN_2C_{13})_8$ | $ClN_2C_{13}$ | 9.16 | 9.74 | 26.63 | 90.00 | 90.00 | 90.00 | Pbca |
| 183L | 0.060 | $(O_2N_5C_{20})_2$ | $O_2N_5C_{20}$ | 5.50 | 10.29 | 15.81 | 87.68 | 88.89 | 79.39 | P-1 |
| 184 | 0.053 | $(N_2C_{20})_4$ | $N_2C_{20}$ | 5.95 | 16.89 | 15.09 | 90.00 | 90.66 | 90.00 | $P2_1/c$ |
| 185 | 0.16 | $(CuCl_4NaN_2C_{13})_4$ | $(CuCl_4NaN_2C_{13})_2$ | 7.59 | 7.00 | 30.47 | 90.00 | 90.91 | 90.00 | $P2_1$ |
| 186 | 0.097 | $(O_5N_2C_{22})_4$ | $O_5N_2C_{22}$ | 11.52 | 7.42 | 22.24 | 90.00 | 91.81 | 90.00 | $P2_1/n$ |
| 187 | 0.041 | $(O_2N_5C_{20})_2$ | $O_2N_5C_{20}$ | 5.50 | 10.29 | 15.80 | 87.65 | 88.93 | 79.35 | P-1 |
| 188 | 0.035 | $(BrOC_{52})_2$ | $BrOC_{52}$ | 12.97 | 13.06 | 13.42 | 70.18 | 82.71 | 60.50 | P-1 |
| 189 | 0.044 | $(Cl_4O_2C_{28})_4$ | $Cl_4O_2C_{28}$ | 17.99 | 7.40 | 18.48 | 90.00 | 118.24 | 90.00 | $P2_1/n$ |
| 190 | 0.039 | $(O_4N_2C_{16})_4$ | $O_4N_2C_{16}$ | 11.58 | 7.49 | 16.72 | 90.00 | 94.19 | 90.00 | $P2_1/c$ |
| 191L | 0.065 | $(BrO_3N_3C_{15})_2$ | $BrO_3N_3C_{15}$ | 16.11 | 5.80 | 7.43 | 90.00 | 94.74 | 90.00 | Pc |
| 192L | 0.030 | $(O_5C_{12})_4$ | $O_5C_{12}$ | 8.88 | 18.17 | 7.72 | 90.00 | 106.04 | 90.00 | Cc |
| 193L | 0.035 | $(O_3N_4C_{27})_4$ | $O_3N_4C_{27}$ | 17.33 | 17.34 | 7.78 | 90.00 | 90.00 | 90.00 | $Pna2_1$ |
| 194L | 0.030 | $(S_2O_3C_{13})_4$ | $S_2O_3C_{13}$ | 11.43 | 7.93 | 16.78 | 90.00 | 108.78 | 90.00 | $P2_1/c$ |
| 195L | 0.097 | $(O_3N_2C_{21})_4$ | $O_3N_2C_{21}$ | 9.56 | 16.03 | 11.84 | 90.00 | 90.00 | 90.00 | $Pca2_1$ |
| 196L | 0.038 | $(ON_3C_{22})_4$ | $ON_3C_{22}$ | 14.22 | 5.58 | 22.06 | 90.00 | 93.44 | 90.00 | $P2_1/n$ |
| 197L | 0.045 | $(S_2ON_5C_{18})_4$ | $S_2ON_5C_{18}$ | 14.77 | 7.35 | 17.98 | 90.00 | 106.96 | 90.00 | $P2_1/n$ |
| 198L | 0.10 | $(ON_3C_{22})_4$ | $(ON_3C_{22})_2$ | 9.62 | 30.72 | 5.86 | 90.00 | 93.68 | 90.00 | $P2_1/c$ |
| 199L | 0.033 | $(ON_3C_{22})_4$ | $ON_3C_{22}$ | 14.23 | 5.59 | 22.07 | 90.00 | 93.90 | 90.00 | $P2_1/n$ |
| 200L | 0.056 | $(O_2N_4C_{15})_2$ | $O_2N_4C_{15}$ | 16.12 | 5.80 | 7.43 | 90.00 | 94.75 | 90.00 | Pc |
| 201L | 0.047 | $(O_5N_4C_{14})_4$ | $O_5N_4C_{13}$ | 5.52 | 17.79 | 15.69 | 90.00 | 91.62 | 90.00 | $P2_1/n$ |
| 202L | 0.051 | $(O_2N_5C_{13})_4$ | $O_2N_5C_{13}$ | 9.34 | 9.28 | 16.52 | 90.00 | 97.74 | 90.00 | $P2_1/n$ |
| 203L | 0.038 | $(O_5N_4C_{14})_4$ | $O_5N_4C_{14}$ | 5.52 | 17.79 | 15.70 | 90.00 | 91.60 | 90.00 | $P2_1/n$ |
| 204 | 0.075 | $(O_2N_2C_{15})_2$ | $O_2N_2C_{15}$ | 7.53 | 8.56 | 10.74 | 75.50 | 84.28 | 84.83 | P-1 |
| 205 | 0.045 | (O4N4C16)$_4$ | O4N4C16 | 5.52 | 17.78 | 15.69 | 90.00 | 91.60 | 90.00 | $P2_1/n$ |
| 206L | 0.043 | $(O_5N_4C_{14})_4$ | $AgCl_2Si_2OC_{36}$ | 19.55 | 15.89 | 20.15 | 90.00 | 90.00 | 90.00 | Pbca |
| 207L | 0.061 | $(S_2N_2C_{26})_4$ | $S_2N_2C_{26}$ | 15.75 | 9.37 | 15.77 | 90.00 | 113.35 | 90.00 | $P2_1/n$ |
| 208L | 0.047 | $(ON_3C_{11})_4$ | $ON_3C_{11}$ | 10.42 | 5.31 | 16.06 | 90.00 | 95.26 | 90.00 | $P2_1/n$ |
| 209L | 0.041 | $(Cl_2O_3N_2C_{17})_2$ | $Cl_2O_3N_2C_{17}$ | 7.49 | 9.13 | 12.43 | 92.37 | 91.85 | 99.09 | P-1 |
| 210L | 0.038 | $(O_4N_2C_8)_4$ | $O_4N_2C_8$ | 6.91 | 13.10 | 10.45 | 90.00 | 106.79 | 90.00 | $P2_1/c$ |
| 211L | 0.043 | $(ON_4C_{20})_4$ | $ON_4C_{20}$ | 7.99 | 13.45 | 14.45 | 90.00 | 93.30 | 90.00 | $P2_1/n$ |
| 212L | 0.036 | $(Cl_2O_3N_2C_{17})_2$ | $Cl_2O_3N_2C_{17}$ | 7.50 | 9.14 | 12.43 | 92.33 | 91.92 | 99.06 | P-1 |
| 213L | 0.037 | $(ClO_3N_4C_{22})_2$ | $ClO_3N_4C_{22}$ | 8.47 | 11.08 | 12.15 | 91.89 | 103.97 | 111.48 | P-1 |
| 214L | 0.072 | $(O_2N_5C_{23})_{16}$ | $O_2N_5C_{23}$ | 26.08 | 26.08 | 16.04 | 90.00 | 90.00 | 90.00 | $I4_1/a$ |
| 215L | 0.086 | $(ON_5C_{17})_4$ | $(ON_5C_{17})_2$ | 8.91 | 12.80 | 13.86 | 94.25 | 98.37 | 101.40 | P-1 |
| 216L | 0.028 | $(ON_4C_{21})_4$ | $ON_4C_{21}$ | 32.00 | 9.25 | 6.46 | 90.00 | 90.00 | 90.00 | $Pna2_1$ |
| 217L | 0.046 | $(O_4N_6C_{20})_4$ | $O_4N_6C_{20}$ | 9.16 | 15.22 | 14.66 | 90.00 | 96.26 | 90.00 | $P2_1/n$ |
| 218L | 0.051 | (O4N5C$_{22}$)$_2$ | O4N5C$_{22}$ | 8.38 | 10.65 | 12.96 | 79.16 | 83.90 | 67.81 | P-1 |
| 219L | 0.046 | $(O_6N_5C_{25})_2$ | $O_6N_5C_{25}$ | 9.18 | 12.03 | 12.06 | 82.37 | 87.42 | 79.50 | P-1 |
| 220L | 0.090 | $(O_3N_5C_{18})_4$ | $O_3N_5C_{18}$ | 16.21 | 16.21 | 14.30 | 90.00 | 90.00 | 90.00 | $P4_2/n$ |

| Sample | R | Cell content | Asymmetric unit | a(Å) | b(Å) | c(Å) | α(°) | β(°) | γ(°) | space group |
|---|---|---|---|---|---|---|---|---|---|---|
| 221L | 0.048 | $(O_3N_3C_{16})_2$ | $O_3N_3C_{16}$ | 9.14 | 9.49 | 10.01 | 79.55 | 63.79 | 83.05 | P-1 |
| 222L | 0.025 | $(ClO_3N_4C_{21})_4$ | $ClO_3N_4C_{21}$ | 7.53 | 15.97 | 16.54 | 90.00 | 90.00 | 90.00 | $P2_12_12_1$ |
| 223L | 0.054 | $(ON_5C_{19})_4$ | $ON_5C_{19}$ | 9.11 | 13.89 | 13.46 | 90.00 | 103.91 | 90.00 | $P2_1/c$ |
| 224L | 0.048 | $(O_2N_5C_{25})_4$ | $O_2N_5C_{25}$ | 11.31 | 9.46 | 21.56 | 90.00 | 94.27 | 90.00 | $P2_1/n$ |

### S10. Availability of the Python source codes

The Python source codes used in this work can be accessed in GitHub:

xzhang222/sR1-method: Python codes for single-atom R1 method.

(https://github.com/xzhang222/sR1-method)

For questions about the sR1 method and the Python codes please contact the author:

xzhang2@tulane.edu

### S11. How to verify the effectiveness of the sR1 method

Though in the body of the paper a model comparison criterion has been suggested to verify that the sR1 calculation can reach approximately the same model as *SHELXT* (Sheldrick, 2015a) does when treating a same dataset, such verification is only auxiliary. In fact, there is independent way to verify the effectiveness of an sR1 result: visually inspecting the resulting model one can immediately tell if the structure has been revealed or has not been revealed at all. Such qualitative visual inspection is performed in daily work. However, it should be acknowledged that one reviewer has raised concern on the usage of such qualitative criterion:

> Qualitative criterion (for example, visual recognition of molecular geometry) is not a scientific method: the molecule could be shifted compared to the real position, or could be differently oriented, or explodes when subjected to mnq. What would the user think when such case occurs?

It should be clarified that for the sR1 method this concern is unnecessary. Whole model shift is immaterial, which results in the same solution. In fact, in sR1 calculation such shift happens all the time. The sR1 method also cannot distinguish between a solution and an inverted solution. Furthermore, the sR1 method is carried out in the P1 space group. The mnq of P1 only contains an identity operator. Upon identity operation, no "explode" is expected. Thus, the only remaining concern is a model that is incorrectly oriented. If a structure is solved by fitting a known model into an unknown structure, this concern is legitimate. However, the sR1 develops a solution in an atom-by-atom manner. A resulting model happens to be chemically recognizable but incorrectly oriented is unlikely to happen. This is because the initial partial structure guides new atoms into "correct" positions and/or atoms of “more incorrectly positioned” ones are gradually replaced by “more correctly positioned” ones. Correct relative positioning and chemical recognizability are developed together. They are two sides of the same coin. Furthermore, the method is centered on minimizing R1 value. The final model corresponds to minimum R1. This means the orientation of the resulting model is close to the best that can match the reflection data. With all these considerations, the chemical recognizability that has been developed in such a process is “the” sign that the correct model has been revealed. Therefore, for the sR1 method, qualitative visual inspection is valid evidence for verifying the effectiveness of the method.

It should be acknowledged that often an R-value is reported. The main factor affecting an R-value is the quality of the dataset. Because the reliability of the resulting model does depend on the quality of the dataset, it is appropriate to report an R-value. However, an R-value alone cannot tell if a structure has been revealed or not. One still needs to visually inspect the model to be certain whether the structure has indeed been revealed or has not been revealed at all. It is obvious that no matter how low an R-value is, if no chemically recognizable model has emerged, no one would admit that the structure has been “solved”. On the other hand, if a chemically recognizable model has emerged, no matter how high the R-value is, at least the structure has been qualitatively revealed. Once it is certain that a structure has been revealed, an R-value can be cited to indicate the quality and reliability of the resulting model.

Conclusion: Visual inspection and only visual inspection can tell whether a structure has been revealed or not. Once it is certain that a structure has been revealed, an R-value can be cited to indicate the quality and reliability of the model. The purpose of the sR1 method is to reveal the structure. For this, only visual inspection can help to determine if a structure has been revealed or has not been revealed.

## References

Sheldrick, G. M. (2015a). *Acta Cryst.* A**71**, 3-8.

Sheldrick, G. M. (2015b). *Acta Cryst.* C**71**, 3-8.

Spek, A.L. (2020). *Acta Cryst.* E**76**, 1-11.

Zhang, X. & Donahue, J. P. (2024). *Acta Cryst.* A**80**, 237-248.